\documentclass[12pt]{article}

\newenvironment{wileykeywords}{\textsf{Keywords:}\hspace{\stretch{1}}}{\hspace{\stretch{1}}\rule{1ex}{1ex}}

\usepackage{amsmath,amssymb}
\usepackage{graphicx}% Include figure files
\usepackage{color}% Include colors for document elements
\usepackage{dcolumn}% Align table columns on decimal point
\usepackage{bm}% bold math
\usepackage[numbers,super,comma,sort&compress]{natbib}
\usepackage{indentfirst}

\definecolor{background-color}{gray}{0.98}

\title{GALAMOST: GPU-accelerated large-scale molecular simulation toolkit}

\author{You-Liang Zhu\footnotemark[1], Hong Liu\footnotemark[1], Zhan-Wei Li\footnotemark[2], Hu-Jun Qian\footnotemark[1], Giuseppe Milano\footnotemark[3],\\ and Zhong-Yuan Lu\footnotemark[1]}

\footnotetext[1]{State Key Laboratory of Theoretical and Computational Chemistry, Institute of Theoretical Chemistry, Jilin University, Changchun 130023, China \\ E-mail: luzhy@jlu.edu.cn}
\footnotetext[2]{State Key Laboratory of Polymer Physics and Chemistry, Changchun
Institute of Applied Chemistry, Chinese Academy of Sciences, Changchun 130022, China}
\footnotetext[3]{Dipartimento di Chimica e Biologia and NANOMATES, Research Centre for NANOMAterials and nanoTEchnology at Universit\`{a} di Salerno, I-84084 via Ponte don Melillo Fisciano (SA), Italy}

\begin{document}

\maketitle

\begin{abstract}
GALAMOST (GPU-accelerated large-scale molecular simulation toolkit) is a molecular simulation package designed to utilize the computational power of graphics processing units (GPUs). Besides the common features of molecular dynamics packages, it is developed specially for the studies of self-assembly, phase transition, and other properties of polymeric systems at mesoscopic scale by employing some lately developed simulation techniques. To accelerate the simulations, GALAMOST contains a hybrid particle-field molecular dynamics technique where particle-particle interactions are replaced by interactions of particles with density fields. Moreover, the numerical potential obtained by bottom-up coarse-graining methods can be implemented in simulations with GALAMOST. By combining these force fields and particle-density coupling method in GALAMOST, the simulations for polymers can be performed with very large system sizes over long simulation time. In addition, GALAMOST encompasses two specific models, i.e., a soft anisotropic particle model and a chain-growth polymerization model, by which the hierarchical self-assembly of soft anisotropic particles and the problems related to polymerization can be studied, respectively. The optimized algorithms implemented on the GPU, package characteristics, and benchmarks of GALAMOST are reported in detail.
\end{abstract}

\begin{wileykeywords}
polymers, molecular dynamics, GPU, anisotropic particles, polymerization.
%A list of five key words or phrases which best characterize the paper are required for indexing.
\end{wileykeywords}

\clearpage

%\section*{\sffamily \Large SUMMARY}

%A short summary of the main contributions in the paper is required. This summary should be carefully prepared for it is automatically the source for most abstracts.

\section*{\sffamily \Large INTRODUCTION} % Not needed for rapid communications

As the computational power continuously increases, molecular dynamics (MD) method has become a more and more common tool in the studies of polymeric and biological systems~\cite{Frenkel2002}. However, some bottlenecks of atomistic MD simulations have restricted their applications on solving the problems related to mesoscopic time and length scales. Recently, many new simulation techniques such as coarse-graining~\cite{Reith2003,Ruhle2009} and enhanced sampling~\cite{Mitsutake2001} were proposed to accelerate simulations and meanwhile maintain the computational accuracy. Nevertheless, developing highly-optimized and parallelized MD programs is also essential to access larger spatial and longer temporal scales. For example, the NAMD~\cite{Phillips2005} program allowed ten-microsecond all-atom MD simulations of protein folding~\cite{Freddolino2008} or the simulations of a complete satellite tobacco mosaic virus with up to 1 million atoms~\cite{Freddolino2006}.
\par
Recently, graphics processing units (GPUs) that are originally developed for rendering images or real-time effects in computer games, have been optimized for intensive computational tasks, certainly including MD simulations. For example, the processor chip of latest generation of NVIDIA GPUs, Tesla K20X (1 Kepler GK110) based on Kepler computing architecture provides a theoretical peak 3.95 Teraflops of single precision compution throughput~\cite{nvidia}. Due to the tremendous computational performance at a fraction of the cost and power consumption as compared with CPUs, GPUs are more and more taken as main processors in MD simulations and supported by popular MD packages, such as DL-POLY~\cite{Smith1996}, AMBER~\cite{Goetz2012}, LAMMPS~\cite{Plimpton1995}, NAMD~\cite{Phillips2005}, and GROMACS~\cite{Berendsen1995}.
\par
The programming model supplied by Compute Unified Device Architecture (CUDA) provides an easier way to harness the computational power of GPUs. For an MD algorithm to be executed efficiently on the GPU, it must be casted into a data-parallel form with extremely massive independent threads of execution in SIMD (single-instruction-multiple-data). The number of threads in a kernel function is scalable and depends on specific algorithm. Due to the different architecture of the GPU from CPU, an existed CPU algorithm of MD has to be redesigned and optimized for the implementation on the GPU. In addition, there are very few of highly optimized GPU-based MD packages designed for simulating polymeric systems efficiently. Thereby, we develop this versatile toolkit, the GALAMOST (GPU-Accelerated LArge-scale MOlecular Simulation Toolkit), for facilitating the studies of various polymeric systems.

\section*{\sffamily \Large A SKETCH OF GALAMOST}

%((Place Computational Methods here. Not needed for review articles))

%((Computational results should be prepared following the IUPAC guidelines (See Journal of Computational Chemistry, 20: 1587-1590 and 20:1591-1592). In particular, it is required that the level of theory employed is appropriate to the problem at hand, and that the sufficient details about methodology are provided to allow the work to be reproduced.)

%((In full papers, this section appears immediately after the introduction. In Rapid Communications, this section appears just before the Acknowledgments.))
Compared with complex biomolecules, polymers with repeated units can be easily modeled in coarse-grained MD simulations, usually without considering long-range electrostatic interactions. The coarse-grained MD has become a powerful tool, accounting for the problems related to self-assembly, phase separation, and other phenomena of polymeric systems. In this paper, we introduce some lately developed coarse-grained simulation methods especially for the studies of polymers, and their optimized algorithms implemented on the GPU. These methods are included in GALAMOST that is highly optimized to run on GPUs with CUDA.
\par
GALAMOST takes a similar strategy of HOOMD-blue~\cite{Anderson2008,hoomd} that is to optimize algorithms for running very efficiently on a single GPU. We have learnt a few techniques from HOOMD-blue and employed them in our package, such as, the SFCPACK~\cite{Anderson2008} sorting method which rearranges the order of data of particles stored in host and device memories to increase the probability of cache hit at memory reading. Different from HOOMD-blue, GALAMOST is purely designed for GPUs and cannot be run on  CPUs independently. The GPU code of GALAMOST can be compiled in single precision or in double precision, but all benchmarks in this work are associated with single precision. The main programming languages are C++ and CUDA C. The compiling configurations and processes are controlled by a Shell Script and a Makefile in Linux operating system, respectively.
\par
\subsection*{\sffamily \large The force fields}
The choices for general force field functions in GALAMOST will be listed in the following sections. In addition, a hybrid particle-field molecular dynamics technique (MD-SCF)~\cite{Milano2009,Zhao2012} for calculating intermolecular interactions has been incorporated in GALAMOST. This technique is similar to Molecular Dynamics the Single-Chain in-Mean-Field scheme developed by M\"{u}ller and co-workers~\cite{Mueller2011}. In this technique, the most computational time-consuming parts in MD, i.e., the intermolecular pair interactions, are replaced by interactions of particles with density fields. It will largely speed up some slowly evolving collective processes in MD simulations, such as micro-phase separation and self-assembly of polymeric systems. In addition to the character of intrinsic speed-up of intermolecular interactions, the algorithms of MD-SCF are very suitable to be parallelized, not only on multi-CPUs, but also on GPUs.
\par
In addition to analytical potentials, numerical potential can be used in GALAMOST by reading the potential table derived from iterative Boltzmann inversion (IBI) method~\cite{Reith2003,Chen2006} or other structure-based coarse-graining methods~\cite{Lyubartsev1995,Harmandaris2007}. The IBI method developed by M\"{u}ller-Plathe and coworkers can successfully generate the coarse-grained numerical potentials in a simpler way~\cite{Reith2003,Karimi-Varzaneh2011}. It derives the coarse-grained potentials by mapping the structural distributions onto the ones obtained either from atomistic simulations or from experiments. With this bottom-up coarse-graining scheme, the derived coarse-grained numerical potentials can be applied in larger systems under the same thermodynamic conditions. Therefore, a framework of multiscale modeling of polymeric systems can be constructed. By further utilizing the computational power of GPUs, the temporal and spatial scales in the simulations of polymers are greatly enlarged again.
\subsection*{\sffamily \large The characteristic models}
Besides some basic functions of general MD, such as coarse-grained MD (CGMD)~\cite{Monticelli2008}, Brownian dynamics (BD)~\cite{Allen1989}, and dissipative particle dynamics (DPD)~\cite{Groot1997}, GALAMOST also includes two tailor-made modules: A soft anisotropic particle model has been incorporated for modeling some kinds of anisotropic particles, and a stochastic chain-growth polymerization reaction model has been developed specially for the studies related to polymerization. By employing these advanced simulation techniques on the GPU, GALAMOST enriches the routes for researchers to investigate polymeric systems via computer simulations. More information about these models is introduced in the following.
\par
To simulate hierarchically self-assembled superstructures of anisotropic particles, a novel anisotropic particle model was proposed by Li et al.~\cite{Li2008}. To be specific, the packing and the self-assembly of disk-like, rod-like, Janus, and triblock Janus particles can be studied by using different anisotropic potentials~\cite{Li2010,Li2011,Li2012,Li2013}. Due to the unique feature of being noncentrosymmetric, the self-assembly patterns of anisotropic particles show great diversity~\cite{Glotzer2004, Zhang2005}. Compared with popular methods which collect a set of spherical particles to be an anisotropic rigid body and keep the rigid body by SHAKE algorithm or Quaternion scheme~\cite{Ryckaert1977, Allen1989, Nguyen2011}, this single-site anisotropic particle model has great superiority on computational efficiency as a result of its simplicity. With the formidable computational power of GPUs, it has great potential in the future studies of plentiful self-assembled superstructures and other properties of anisotropic particles.
\par
To study the phenomena related to polymerization, a stochastic chain-growth polymerization model coupled with CGMD was proposed recently by Liu et al.~\cite{Liu2007}. This model has its applications in, for example, polymerization-induced phase separation, surface-initiated polymerization, and so on~\cite{Liu2007,Liu2009,Xue2010,Liu2012}. Surface-initiated polymerization is not trivial since it is of great relevance with the performance of the designed materials of polymer brushes~\cite{Turgman-Cohen2011,Turgman-Cohen2012,Jalili2012}. The properties of polymer brushes fabricated via surface-initiated polymerization are synergistically controlled by excluded volume effect, diffusion of monomers, and polymerization~\cite{Liu2009}. The implementation of the stochastic chain-growth polymerization model in GALAMOST supplies a powerful tool to deal with these types of problems~\cite{Liu2007}.

\section*{\sffamily \Large MOLECULAR DYNAMICS METHODS AND ALGORITHMS}
To make the operations in the code of GALAMOST clear enough, the device memory reads and writes are denoted by $\Leftarrow$ and $\Rightarrow$ in pseudocodes, respectively. The $\leftarrow$ denotes the operations from local register to local register. In addition, the value in [ ] is used to denote an index of array and ( ) denotes a specified calculation. To cater to the data-parallel calculation mode of GPU, the neighbor list (NL) array which records the tags of all neighboring particles of each particle and the cell list (CL) array which records the information (i.e., tag, type, and coordinate vector) of the particles in each cell, are stored in device memory in table-like form, as illustrated in Fig.~\ref{Fig:fig1}. The neighbor number (NN) array records the number of the neighboring particles of each particle in neighbor list. The cell particle number (CN) array records the number of particles in each cell. The bond list (BL) array in device memory which records the tags of the bond-connected particles for each particle takes the same storage form as NL, as illustrated in Fig.~\ref{Fig:fig2}. The bond number (BN) array records the number of the bonds of each particle. Since one dimensional array in device memory is the most convenient to be allocated, copied, and used, we transform the  multi-dimensional array into one-dimensional array form enhancing the computational efficiency. For example, in the operation of reading neighboring particle tag $j$$\Leftarrow$NL[$(i, k)$] in Algorithm 2, the $(i, k)$ denotes the transformation ($k \ast Np+ i$) of an index of NL array where Np is the total number of particles in the system, $i$ is the particle tag which also denotes the horizontal index of the table of NL, and $k$ is the vertical index of the table. This operation means reading the tag of the $k$-th neighboring particle of particle $i$ from NL. Besides the coordinate vector array $\vec{R}$, velocity vector array $\vec{V}$, force vector array $\vec{F}$, and type of particles array T, other arrays and variables are defined in each algorithm. The description of each method and its optimized algorithm implemented on the GPU are introduced in the following.

\subsection*{\sffamily \large Short range non-bonded forces}
In polymer systems, the net non-bonded force of each particle is produced by summing all the non-bonded forces of neighboring particles. The standard algorithm is to employ an NL that lists the interacting particles for each particle, built beforehand. Because of the independence of parallel CUDA threads, a pair of interacting particles is inevitably included independently in NL in the mode that one thread calculates and sums all non-bonded forces of a particle. The employed GPU algorithm of NL that is built after binning all particles to CL is similar to the practice~\cite{Anderson2008}. The processes of building CL and NL are both achieved by GPU calculations. In the kernel function of building CL, a thread corresponds to a particle, and then this thread calculates the index of the cell in which the particle resides and adds its tag together with its coordinate vector into the cell list. Its algorithm is almost the same as Kernel 1 in Algorithm 1, except that the tag instead of the type is combined with the coordinate vector. Based on the CL that has been built, the NL for each particle can be easily built by searching neighboring particles from the surrounding 27 cells (including the cell that the target particle resides in). The NL records all neighboring particles of each particle within a distance that is the sum of the cutoff of non-bonded interactions, $r_{cut}$, and the buffer distance, $r_{buffer}$. The NL only needs to be updated when any particle has moved a distance more than half of the buffer distance. The non-bonded forces for each particle can be calculated and summed by a corresponding thread on the  GPU by searching the particles within $r_{cut}$ from NL. The common non-bonded potential energy functions are included in GALAMOST, which are
Lennard-Jones (LJ):
\begin{eqnarray}
U_{\mathrm{LJ}}(r_{ij})=4\epsilon\left[\left(\frac{\sigma}{r_{ij}}\right)^{12}-\left(\frac{\sigma}{r_{ij}}\right)^{6}\right],
\end{eqnarray}
where $\epsilon$ is the depth of the potential well, $\sigma$ is the finite distance at which the interparticle potential is zero, and $r_{ij}$ is the distance between the particles;\\
harmonic repulsion:
\begin{eqnarray}
U_{\mathrm{harmonic}}(r_{ij})=\alpha\left(1-\frac{r_{ij}}{r_{cut}}\right)^{2},
\end{eqnarray}
where $\alpha$ and $r_{cut}$ set the maximum energy penalty and maximum interaction distance, respectively;\\
Gaussian repulsion:
\begin{eqnarray}
U_{\mathrm{Gaussian}}(r_{ij}) = \varepsilon \exp \left[ -\frac{1}{2}\left( \frac{r_{ij}}{\sigma} \right)^2 \right],
\end{eqnarray}
where $\varepsilon$ and $\sigma$ determine the energy and length scales, respectively.
%\textcolor[rgb]{1.00,0.00,0.00}{
%and normal electrostatic Coulombic potential:
%\begin{eqnarray}
%U_{\mathrm{el}}(r_{ij}) =\frac{q_{i}q_{j}}{4\pi\epsilon_{0}\epsilon_{r}r_{ij}},
%\end{eqnarray}
%where $q$, $\epsilon_{0}$, and $\epsilon_{r}$ are charges, electric permittivity of free space, and dielectric constant, respectively.
%}
\subsection*{\sffamily \large The MD-SCF treatment of interparticle forces}
In the framework of SCF theory, a molecule is regarded to be interacting with the surrounding molecules not directly but through a density field. The interactions on particles derived from the minimization of the free energy of density fields efficiently push forward the evolution of the configuration of the system. However, all intramolecular interaction terms (bond, angle and other possible intramolecular interactions) are usually considered as the regular forms in MD. The different scales of interactions can be mitigated by treating the stiff intramolecular interactions by propagating the system configuration via a small time step, but treating soft intermolecular interactions by updating the density field derived from the coordinates of particles only after many of these small time steps.
\par
According to the spirit of SCF theory, the main issue is to derive the partition function of a single molecule in an external potential $V(\mathbf{r})$ and further to obtain a suitable expression of the $V(\mathbf{r})$ and its derivatives. Starting from the partition function, the functional form of the interaction term and using the saddle point approximation, the external potential has the form:~\cite{Milano2009}
\begin{eqnarray}
V_{K}(\mathbf{r})=k_{B}T\sum_{K^{'}}\chi_{KK^{'}}\phi_{K^{'}}(\mathbf{r})+\frac{1}{\kappa}(\sum_{K}\phi_{K}(\mathbf{r})-1),
\end{eqnarray}
where each component is specified by an index $K$, $\kappa$ is the compressibility, $\chi$ represents the mean field parameter which is related to the Flory$-$Huggins parameter, and $\phi$ is the density distribution.
In the case of a mixture of two components A and B, the mean field potential acting on a particle of type A at position $\mathbf{r}$ is given by
\begin{eqnarray}
V_{A}(\mathbf{r})=k_{B}T[\chi_{AA}\phi_{A}(\mathbf{r})+\chi_{AB}\phi_{B}(\mathbf{r})]+\frac{1}{\kappa}(\phi_{A}(\mathbf{r})+\phi_{B}(\mathbf{r})-1).
\end{eqnarray}
Then the force acting on particle A at position $\mathbf{r}$ imposed by the interaction with the density field is
\begin{eqnarray}
F_{A}(\mathbf{r})=-\frac{\partial V_{A}(\mathbf{r})}{\partial\mathbf{r}}=-k_{B}T(\chi_{AA}\frac{\partial\phi_{A}(\mathbf{r})}{\partial\mathbf{r}}+\chi_{AB}\frac{\partial\phi_{B}(\mathbf{r})}{\partial\mathbf{r}})-\frac{1}{\kappa}(\frac{\partial\phi_{A}(\mathbf{r})}{\partial\mathbf{r}}+\frac{\partial\phi_{B}(\mathbf{r})}{\partial\mathbf{r}}).
\end{eqnarray}
\par
The algorithm on the GPU about the calculation of the forces acting on particles from density fields is different from that on the CPU in many places. For example, assigning and adding the density fractions of each particle to the eight vertices of the cell in which the particle resides, are implemented simply one by one in serial CPU code. However, in GPUs, one thread will take charge of one particle in kernel function and assign the density fractions of corresponding particle to the eight vertices of the cell in which the particle resides. In this process, atomic function from CUDA is inevitably employed to compel more than one thread to queue, so that these threads will add density fractions to the same vertex sequentially in order to avoid interference with  each other. In practice, we employ another algorithm illustrated in Part 1 of Algorithm 1 which seems a little complex, but more efficient. First, we build a CL which records the coordinate vectors and types of the particles in each cell. And then, one thread takes charge of one vertex and collects the density fractions of the particles from the eight cells near this vertex.
\par
The algorithms for the update of density gradients (density derivatives) from density distributions and the calculation of the forces from density gradients are explained in two kernels in Part 2 of Algorithm 1. In addition, a trick has been employed to accelerate the most time-consuming function Kernel 4. A composition list which records the types of the components with non-zero density derivatives is used. It can reduce the \texttt{\textbf{for}} loops significantly in Kernel 4 for multi-component systems. This trick works well especially for self-assembly systems in dilute solution. The speed-up by this trick depends on the number of components in the system. Normally we have 5\%-40\% speed-up of overall performance for multi-component systems.

\subsection*{\sffamily \large Bonded forces}
The bonded forces generally include the bond, angle, and torsion forces. Because of the execution mode of SIMD in GPU, it is efficient that a thread calculates and sums the bond, angle, and torsion forces of a particle in the kernel functions. Thus, how to store the topological information of the molecules is of great relevance. As illustrated in Fig.~\ref{Fig:fig2}, we need two arrays, BL and BN to record all bonds. The tags in a column of BL table indicate the particles which are bonded to the target particle, and correspond to the horizontal index of the BL table. Therefore, we can employ the thread whose number is equal to the number of particle to calculate the bond forces of each particle independently. For angles and torsions, we store them with the same form as bonds, except that the position (side or middle) of the target particle in an angle or a dihedral needs to be indicated. Although the force of a bond will be computed twice (three times for an angle and four times for a torsion) on device according to the form of the storage, it casts the computation well into parallel mode for separated threads of GPUs and is efficient for data copy between host and device memories. The common bonded potential energy functions in GALAMOST are
%\begin{eqnarray}
%U_{bond}(r_{ij}) =  = \left\{\begin{array}{ll} k^{bond}(r_{ij}-r_{0})^{2},\\
%k^{bond}(r_{ij}-r_{0})^{2},.
%\end{array} \right.\,
%\end{eqnarray}
\begin{eqnarray}
U_{bond}(r_{ij})=k^{bond}(r_{ij}-r_{0})^{2},
\end{eqnarray}
in which $r_{ij}$ is the instantaneous length of the bond, $r_{0}$ is the equilibrium length of the bond, and $k^{bond}$ is the spring constant;
\begin{eqnarray}
U_{angle}(\theta_{ijk})=k^{angle}(\theta_{ijk}-\theta_{0})^{2}
\end{eqnarray}
%\textcolor[rgb]{1.00,0.00,0.00}{
or alternatively
\begin{eqnarray}
U_{angle}(\theta_{ijk})=k^{angle}\{\mbox{cos}(\theta_{ijk})-\mbox{cos}(\theta_{0})\}^{2},
\end{eqnarray}
%}
in which $\theta_{ijk}$ is the angle in radians between vectors $\vec{r}_{ij}$ and $\vec{r}_{jk}$, $\theta_{0}$ is the equilibrium angle, and $k^{angle}$ is the angle force constant;
\begin{eqnarray}
U_{dihedral}(\phi_{ijkl})= k^{dihedral}\left[ 1 + \cos\left(n  \phi_{ijkl} - \delta \right) \right],
\end{eqnarray}
in which $\phi_{ijkl}$ is the angle in radians between the planes $(i,j,k)$ and $(j,k,l)$, the integer constant $n$ indicates the periodicity, $\delta$ is the phase shift angle, and $k^{dihedral}$ is the multiplicative constant.
\subsection*{\sffamily \large Numerical forces}
The numerical non-bonded, bond, angle, and torsion potentials can be derived from IBI~\cite{Karimi-Varzaneh2011} or reverse Monte Carlo method~\cite{Lyubartsev1995}. With IBI method, the procedure starts with the potentials of mean force as guessed potentials and then optimizes the potentials iteratively by mapping the structural distributions (i.e., radial distribution function, RDF) onto the ones obtained either from atomistic simulations or from experiments. The resulting numerical potentials usually take the form as a table in which the potential values at discrete grid points of distance are given. In the treatment of tabulated potentials, the initial input potential tables on grid points of $r$ are transformed to the tables (arrays) on grid points of $z = r^{2}$. With this trick, the $r = \mbox{SQRT}(r^{2})$ in the inner loop of force calculation is avoided, and the force is then calculated by
\begin{eqnarray}
F = -\mathbf{r}\frac{\partial V(r)}{\partial r}\frac{1}{r}=-2\mathbf{r}\frac{\partial V(z)}{\partial z}.
\end{eqnarray}
Within each interval between the grid points, potentials are fitted to a cubic spline function~\cite{Forsythe1977}, more specifically, for each $x_{i}<x<x_{i+1}$,
let $\delta=x-x_{i}$, $V(x)$ is represented by:
\begin{eqnarray}
V(x)=c_{0}+c_{1}\delta+c_{2}\delta^{2}+c_{3}\delta^{3},
\end{eqnarray}
where $x$ corresponds to $z$, $\theta$, and $\varphi$ for particle-particle distance square, bending angle, and torsion angle, respectively. $i$ is the index of the grid point and $c_{0}$ is the starting potential value of each grid point. Other parameters $c_{1}$, $c_{2}$, and $c_{3}$ are chosen to make the values of the first derivative and the second derivative at both ends of interval $x_{i}$ and $x_{i+1}$ equal to the correct values of function $V$. In this way, the potential is continuous up to second order through the whole interaction range.
\par
The implementation of this numerical potential method in GALAMOST is illustrated by the example of numerical non-bonded force calculation in Algorithm 2. The texture cache is used in line 11 of Algorithm 2 in the case of numerical potential for random memory access of the parameters $c$ (including $c_{0}$, $c_{1}$, $c_{2}$, and $c_{3}$) from the global memory in device, since the number of grid points is always several thousands and the maximum share memory with 48 KB is not enough to store the array of $c$. In addition, the judgement for whether or not the distance between a pair of particles exceeds the cutoff is done in line 10 of Algorithm 2 by checking if the index of grid point exceeds the maximum value.

\subsection*{\sffamily \large Integration methods}
In GALAMOST, we use velocity-Verlet algorithm~\cite{Frenkel2002} for the integration of the equations of motion for CGMD and BD. Three thermodynamic ensembles can be chosen in GALAMOST, i.e., the microcanonical ensemble (NVE), the canonical ensemble (NVT), and isothermal-isobaric ensemble (NPT). The constant temperature can be controlled via Nos\'{e}-Hoover thermostat~\cite{Hoover1985} or Andersen thermostat~\cite{Andersen1980}, and the constant pressure can be controlled via Andersen barostat~\cite{Andersen1980,Hoover1986,Martyna1994}. Particles can be placed in a cubic or cuboid box with periodic boundary conditions~\cite{Allen1989}.
\par
In addition, a modified velocity-Verlet algorithm suggested by Groot and Warren (GW-VV)~\cite{Groot1997} is used for integrating the equations of motion in DPD. A half-step leapfrog algorithm together with Berendsen thermostat is used for the soft anisotropic particle model~\cite{Li2008}. Because of the independence of the integration of each particle, a thread takes charge of the motion of a particle, and updates its coordinate and velocity vectors in kernel functions in all integration methods.

\section*{\sffamily \Large CHARACTERISTIC COARSE-GRAINED MODELS}
\subsection*{\sffamily \large Soft anisotropic particle model}
By adding two degrees of freedom of rotation, our one-site anisotropic particle model can be used to describe disk-like, rod-like, diblock and triblock Janus particles. We have successfully examined the packing and the self-assembly of anisotropic particles with this model~\cite{Li2008,Li2010,Li2011,Li2012,Li2013}.
\par
In the simulations of disk-like and rod-like particles, we adopt a soft anisotropic potential on the basis of the conservative potential in DPD. It can be expressed as
\begin{eqnarray}
\label{Eq-U1}
U_{ij}=(1-\mu f^{\nu})\frac{\alpha_{ij}}{2}(1-r_{ij})^{2},
\end{eqnarray}
where the magnitude of $\alpha_{ij}$ controls the strength of repulsion, $\mu$ controls the shape of the particles, and $\nu$ controls the angular width of repulsion. The disk-like or rod-like particle can be described by different expressions of anisotropic factor $f$.
In disk-like particle model~\cite{Li2008,Li2010}, the anisotropic factor is
\begin{eqnarray}
f=\frac{(\mathbf{n}_{i}\cdot\mathbf{r}_{ij})(\mathbf{n}_{j}\cdot\mathbf{r}_{ij})}{r_{ij}^{2}},
\end{eqnarray}
where $\mathbf{n}_{i}$ and $\mathbf{n}_{j}$ are unit vectors assigning the orientations to particles $i$ and $j$, respectively.
$\mathbf{r}_{ij} =\mathbf{r}_{i} -\mathbf{r}_{j}$ is the interparticle vector. In rod-like particle model~\cite{Li2011}, the anisotropic factor is \begin{eqnarray}
f = \mbox{sin}\theta_{i}\mbox{sin}\theta_{j}
\end{eqnarray}
where $\theta_{i}$ is the angle between $\mathbf{n}_{i}$ and the interparticle vector $\mathbf{r}_{ji}=-\mathbf{r}_{ij}$, $\theta_{j}$ is
the angle between $\mathbf{n}_{j}$ and $\mathbf{r}_{ij}$. For diblock and triblock Janus particles, we take an anisotropic potential with attractive tail,
\begin{eqnarray}
\label{Eq-U2}
U_{ij}=\frac{\alpha^{R}_{ij}}{2}(1-r_{ij})^{2}-f^{\nu}\frac{\alpha^{A}_{ij}}{2}(r_{ij}-r_{ij}^{2}),
\end{eqnarray}
where the magnitude of $\alpha^{R}_{ij}$ controls the strength of repulsion, $\alpha^{A}_{ij}$ controls the strength of attraction, and $\nu$ controls the angular width of attraction.
In Janus particle model~\cite{Li2012}, the anisotropic factor is
\begin{eqnarray}
f = \left\{\begin{array}{ll} \mbox{cos}\frac{\pi\theta_{i}}{2\beta}\mbox{cos}\frac{\pi\theta_{j}}{2\beta} & \mbox{if cos$\theta_{i}\geq$ cos$\beta$ and cos$\theta_{j} \geq$ cos$\beta$}\\
0 & \mbox{otherwise}.
\end{array} \right.\,
\end{eqnarray}
The size of the attractive patches is described by the Janus balance $\beta$. In triblock Janus particle model~\cite{Li2013}, the anisotropic factor is
\begin{eqnarray}
f = \left\{\begin{array}{ll} \mbox{cos}\frac{\pi\theta^{'}_{i}}{2\beta}\mbox{cos}\frac{\pi\theta^{'}_{j}}{2\beta} & \mbox{if cos$|\theta_{i}|\geq$ cos$\beta$ and cos$|\theta_{j}|\geq$ cos$\beta$}\\
0 & \mbox{otherwise},
\end{array} \right.\,
\end{eqnarray}
where $\theta^{'}_{i} = \mbox{arccos}(|\mbox{cos}\theta_{i}|)$, and $\theta^{'}_{j} = \mbox{arccos}(|\mbox{cos}\theta_{j}|)$. The structures of these four kinds anisotropic particles are illustrated in Fig.~\ref{Fig:fig3}.
\par
The translational displacements of anisotropic particles follow Newton's equations of motion. The equations of rotational motion can be written as
\begin{eqnarray}
\dot{\mathbf{n}}_{i}=\mathbf{u}_{i},
\end{eqnarray}
\begin{eqnarray}
\dot{\mathbf{u}}_{i}=\mathbf{g}_{i}^{\bot}/I+\lambda\mathbf{n}_{i},
\end{eqnarray}
where $I$ is the moment of inertia. $\mathbf{u}_{i}$ is defined as the time derivative of the orientation $\mathbf{n}_{i}$. The quantity $\lambda$ can be taken as a Lagrange multiplier. Physically, $\mathbf{g}_{i}^{\bot}$ corresponds to the perpendicular component of torque, responsible for the rotation of the particle. The optimized algorithms implemented on the GPU for these four types of anisotropic particles have been included in GALAMOST. We illustrate the overall algorithm in Algorithm 3. The expressions of calculation ($\vec{R}_{i}$,$\vec{R}_{j}$,$\vec{n}_{i}$,$\vec{n}_{j}$) in Kernel 2 for translational force and torque are dependent on specific anisotropic potential (details can be found in Refs.~\cite{Li2008,Li2010,Li2011,Li2012,Li2013}). The equations of both translational and rotational motion are integrated via a half-step leapfrog algorithm. With implementation on GPUs, this one-site anisotropic particle model supplies a powerful tool to study the ordered packing and self-assembly of noncentrosymmetric particles.
\subsection*{\sffamily \large Chain-growth polymerization model}
In this model, we consider free radical linear chain-growth polymerization, i.e., $mA\rightarrow(- A -)_{m}$ for monomer A. Polymerization probability $P_{r}$ is set to determine whether a monomer will react with an active end or not in a reaction step and is coupled to the real reaction rate $r_{p}$ by
\begin{eqnarray}
\label{Eq-U}
r_{p}=-\frac{d[M]}{dt}=\frac{[P^{*}]P_{r}}{\tau},
\end{eqnarray}
where $[M]$ is the free monomer concentration, $[P^{*}]$ is the concentration of growth centers, and $\tau$ is the reaction time interval~\cite{Liu2007,Liu2009}. In a time interval, if a polymerization reaction event takes place between a monomer and an active end, a bond connection should be added between them and then the active end should be transferred to the new end. Therefore, how the bond connection relationship is stored in device memory is of great relevance, as illustrated in Fig.~\ref{Fig:fig2}.
\par
We have designed a chain-growth polymerization algorithm on the GPU and illustrate it by pseudocode in Algorithm 4. It should be emphasized that the atomic function atomicMax() in line 12 of Algorithm 4 is employed to handle the situation that more than one active end reacts with the same monomer simultaneously. It will avoid the conflicted operations on the same variable by two or more threads in kernel function. An explanation may be needed for the operation $j$$\Rightarrow$BL[($i$, Num$_{i}$)] in line 17 of Algorithm 4. ($i$, Num$_{i}$) represents the calculation (Num$_{i}$$\ast$Np + $i$) which gives an index of BL array to record the tag of newly connected particle, where $i$ and $j$ are the particle tags, Num$_{i}$ is the number of the stored particles in the column of particle $i$ in BL, and also is the vertical index of the table to record the newly connected particle tag $j$.

\section*{\sffamily \Large PERFORMANCE AND VALIDATION}
%((Place Results here. Not needed for review articles.))
\subsection*{\sffamily \large DPD liquid and LJ liquid}
In most cases, the NL method for short range non-bonded interactions is the most time-consuming part. But NL normally does not require updating every step, thus  equal share of costing time to each step is acceptable and comparable to non-bonded force calculation. The large buffer distance can reduce the frequency of the update of NL. However, the number of the particles recorded in NL for each particle will be increased proportionally to ($r_{cut}$+$r_{buffer}$)$^{3}$, which will result in more costing time of non-bonded force calculation by searching non-bonded interacting particles from a larger range. Thereby, a proper buffer distance exists in each simulated system.
\par
When any particle has moved a distance more than half of the buffer distance, the NL needs to be rebuilt. Thereby the ``softness'' of the particles and the magnitude of integration step, which are related to the speed of migration of particles, influence the performance of GALAMOST greatly. We make benchmarks for GALAMOST by simulating LJ particles and soft DPD particles on GPUs, respectively, and compare the results with HOOMD-blue 0.10.0 at the same conditions of simulations. Steeper interaction profiles  request a necessarily smaller integration step, while lower updating frequency of NL yields a higher performance of GALAMOST. We can draw this conclusion from Fig.~\ref{Fig:fig4}, which shows the results for the LJ liquid systems with a packing fraction of 0.2 at an integration step of 0.001 with $r_{cut}$=3.0, and the DPD liquid systems with a reduce number density of 3.0 at an integration step of 0.04 with $r_{cut}$=1.0. Because the efficient O(N) NL method has been employed, the average costing time per each step is proportional to the system size N (the number of particles). According to the comparisons, the performance of GALAMOST is about 10\% slower than HOOMD-blue in LJ liquid simulations. However, the performance of GALAMOST which also employs the Saru~\cite{Phillips2011} pseudo-random number generator (PRNG) in DPD to generate the stochastic force for each pair of interacting particles is about 30\% faster than HOOMD-blue in DPD liquid simulations.
\par
In GALAMOST, we especially focus on saving the valuable device memory by allocating the only-used arrays in device memory. Thereby, up to 2.2 million LJ liquid particles or 3.0 million DPD liquid particles can be simulated by GALAMOST on GTX 580 with 1.5 gigabytes global device memory. As compared with GALAMOST, at most 1.0 millon LJ liquid particles or 1.5 millon DPD particles can be simulated by HOOMD-blue on the same device at the same simulation conditions. The simulated system size is proportional to the volume of device memory. Accordingly, the Tesla K20X with 6 gigabytes global device memory should effectively enlarge the maximum particle number four folds of the simulated systems by GTX 580. The memory sorting by SFCPACK~\cite{Anderson2008} algorithm which can significantly reduce the costing time of non-bonded force calculation is redesigned in GALAMOST to cope with the systems with fast-moving particles. Only the coordinates of the particles are sorted in GALAMOST. This brings a good performance of GALAMOST on simulating soft particle systems which usually need a high frequency of particle data sorting, such as DPD liquid.
\subsection*{\sffamily \large Simulating phospholipids in water with MD-SCF}
The coarse-grained models for phospholipids and water had been built in the framework of MD-SCF in previous works by Milano et al~\cite{Nicola2011,Nicola2012}. Here, we compare the performance of GALAMOST with parallel CPU code OCCAM~\cite{occam} about the MD-SCF method by simulating the phospholipids in water.
\par
Specifically, the phospholipid dipalmitoylphosphatidylcholine (DPPC) molecule model is used in this benchmark work. To verify the correctness of GALAMOST, we have checked the density profiles of different components. For example, the comparison of electron density distributions along the normal direction of lamellar phase of a system with 208 DPPC and 1600 water molecules at 325 K between GALAMOST on the GPU and OCCAM on the CPU has been done, as illustrated in Fig.~\ref{Fig:fig5}. The distributions are measured in single precision on the GPU and in double precision on the CPU. The difference between them can be regarded in the range of fluctuation. To further validate GALAMOST, we simulate the self-assembled structures of four DPPC and water systems at different water/DPPC ratios. The snapshots of equilibrium structures are shown in Fig.~\ref{Fig:fig6}. The comparison of performances between OCCAM and GALAMOST is given by testing two systems, i.e., a smaller lipid and water (LW) system, LW1, and a larger one, LW2. The results ar shown in Fig.~\ref{Fig:fig7}. As we can see, the performance of GALAMOST in single precision running on a single GPU is far beyond the performance of OCCAM in double precision running on 96 parallel CPUs. The significant speed-up is not only attributed to the powerful computational capability of GPUs, but also benefits from the optimization of the algorithms on the GPU.
\subsection*{\sffamily \large Applying numerical potential to simulate polystyrene melt}
The numerical potential method in GALAMOST is mathematically similar to that in IBIsCO~\cite{Karimi-Varzaneh2011}. We validate the GPU code by comparing the bond and angle distributions of polystyrene melt~\cite{Qian2008} simulated by GALAMOST with that of the same system simulated by IBIsCO. The chain length of  polystyrene is 50 monomers (50 particles in CG model) long. The distributions have been measured in single precision on the GPU and in double precision on the CPU. The comparisons of the results are shown in Fig.~\ref{Fig:fig8}. The bond and angle distributions dN/N$_{total}$ are calculated at the space of 0.01 nm and 1 degree, respectively. As we can see, the differences between the curves of CPU and GPU simulations are very tiny and the two lines in each subfigure can be regarded as overlapping with each other. Furthermore, we have checked the time-dependent property, i.e. the diffusion coefficient of polystyrene chain, obtained in CPU and GPU simulations. There is also no noticeable difference. We further compare the performance of GALAMOST in single precision running on a single GPU with that of IBIsCO in double precision running on 8 parallel CPUs by simulating polystyrene melt with various system sizes. To guide the eyes, we multiply the time steps per second (TPS) of 8 parallel CPUs by 10 and then compare it with the TPS of a single GPU, as shown in Fig.~\ref{Fig:fig9}. It is clear that more particles are simulated, the advantage of GPUs is more obvious.
\subsection*{\sffamily \large Using soft anisotropic particle model to describe triblock Janus particle self-assembly}
The parallel CPU codes for one-site anisotropic particle model had been well established before~\cite{Li2008,Li2010,Li2011}. By porting the algorithms from CPU code to GALAMOST, the performance of this anisotropic particle model has been promoted greatly. We have verified the correctness of  GALAMOST by comparing many simulated quantities with the corresponding CPU code, including transitional and rotational energies, momenta, and temperatures, as well as self-assembled structures. The simulations in Ref.~\cite{Li2013} that had been accomplished by GALAMOST can further show the validation. From disk-like particle model to triblock Janus particle model, the anisotropic potential forms become more and more complex. As a result, the computational efficiency will be lowered by more sophisticated force and torque calculations. Thereby, we select the most computational time-consuming triblock Janus particles as the benchmark system to present the performance. The model details about the triblock Janus particles were introduced in Ref.~\cite{Li2013}. We have simulated these triblock particles with different numbers and show detailed performances in Table~\ref{triblockJanus}. The parameters of all the tested systems are set the same as $\alpha^{R}$=396, $\alpha^{A}$=132, $\beta$=65 and $\nu$ =0.5.
\subsection*{\sffamily \large Surface initiated polymerization}
In the GPU code for the stochastic chain-growth polymerization model, some places of the algorithms have been tuned to fit the computational mode of the GPU, such as handling the situation that more than one active end reacts with the same monomer simultaneously. We have validated GALAMOST by reproducing the same polymer brush structure at certain initiator density and polymerization rate for surface initiated polymerization on a flat surface. The simulations in Ref.~\cite{Liu2012} which have been done by GALAMOST can further give the validation of our GPU code.
\par
Here, we further apply this chain-growth polymerization model to simulate the surface-initiated polymerization from the outer surface of a ball to test the performance. The snapshot of the ball with grafted chains after a period of polymerization is given in Fig.~\ref{Fig:fig10}. In this system, the particles that compose the surface of the ball are frozen: They do not move in the simulations, but have interactions with surrounding particles. The initiators that can induce the polymerization with free monomers are distributed uniformly on the surface of the ball. The specific performances of GALAMOST on the GPU for this system with different number of particles are listed in Table~\ref{polymerization}. The performance data are all recorded after 1 million time steps when the chains are densely grafted to the surfaces by polymerization. In addition, we employ the CUDA profiler to profile the program for polymerization simulation. The costing times of some time-consuming kernel functions and polymerization kernel function are all given in Fig.~\ref{Fig:fig11}. As we can see, polymerization function only  costs a small fraction of simulation time, which can be neglected by comparing to neighbor list construction or non-bonded force computation.
\subsection*{\sffamily \large Precision tests and performances}
The early versions of NVIDIA GPUs did not support double precision (DP) floating-point format since single precision (SP) was enough for graphics rendering. However, scientific algorithms typically require DP format and high standard of floating-point operations. Now the Tesla series of GPU cards from NVIDIA can provide a good performance for DP operations: The DP to SP performance ratio is about 1/2 and thus equivalent to that in CPUs. However, the much cheaper gaming cards (such as GeForce) still only support DP operations at a fraction of the speed of SP operations.
\par
To gain highest performance at lowest cost, we have optimized GALAMOST for SP operations. But GALAMOST also supports CGMD simulations in DP format. We have carefully conducted precision tests and calculated round-off errors in MD simulations with GPU DP, GPU SP, CPU DP and CPU SP operations. The numerical stability of an MD simulation can be reflected by its ability to conserve the total energy in microcanonical ensemble. We therefore focus on a standard benchmark system of LJ liquid with reduced  $\rho= 0.8$ and $T=1.0$ in a cubic simulation box with side length $l=12$. The relative root mean square (RMS) deviations of total energy of LJ liquid in microcanonical ensemble with different integration time steps $dt$ and cutoff distances~\cite{Toxvaerd2012} are calculated by using GALAMOST in GPU DP and SP formats, and by using GROMACS in CPU DP and SP formats, respectively.
All the LJ liquid systems are first equilibrated in canonical ensemble with $dt = 0.0005$ for 5000 time units. Then in microcanonical ensemble simulations, the total energy data are recorded every 0.1 time unit for each trajectory with 1000 time units long. We obtain the average total energy $\bar E$ by averaging these $N = 10000$ energy data and then calculate their RMS deviations using $\Delta E = \sqrt{\frac{1}{N}\sum_{i=1}^{N} (E_i-\bar E)^2}/|\bar E|$. The results are shown in Fig.~\ref{Fig:fig12}.

%Many factors such as the smoothness of potential, magnitude of $dt$ as well as precision arithmetic influence the energy conservation. Thereby, to test numerical stability, we select a simple LJ liquid system with number density , $\epsilon$ = 1, $\sigma$ = 1, temperature $K_{B}T = 1$, and in a cubic box with side length $l = 12$ as testing system. The CPU results were measured on the prestigious package GROMACS with the same set parameters at real unit ($\rho = 0.8/$nm$^{3}$ , $\epsilon$ = 1 KJ/mol, $\sigma$ = 1 nm, $T = 1000/8.314$ K, and $l = 12$ nm).
\par
As we can see, the integration time step $dt$, the cutoff distance, and the numerical precision are all influencing the numerical stability of the MD simulations. In general, the simulation results obtained with either GPU or CPU codes with the same numerical precision are quite similar. For MD simulations with $r_{cut}= 3.4$, the RMS deviations of total energy are overall larger than the results with a larger cutoff distance,  $r_{cut}= 6.0$. This can be attributed to the cutoff noise, i.e., the LJ forces are not continuous at the cutoff distance, which will induce impulses in the simulations. Apparently, increasing the cutoff distance will reduce the noise and then enhance the stability of the MD simulations. The integration time step also has a substantial influence on RMS deviations of total energy. For MD simulations with $r_{cut}= 3.4$, there is no apparent difference in the curves showing the dependence of RMS deviations of total energy on $dt$: For $dt\le 0.004$, the RMS deviations of total energy are all smaller than $4\times 10^{-5}$, while for $dt > 0.004$, $\Delta E$ increases with increasing $dt$. Thus for simulations with comparatively short cutoff distance, there is no systematic difference between GALAMOST with GPU DP and GPU SP and GROMACS with CPU DP and CPU SP operations, because the disturbance of numerical precision on MD is not obvious under the large cutoff noise of LJ forces. For MD simulations with $r_{cut}= 6.0$, the RMS deviations of total energy in simulations with GPU SP and CPU SP operations are several times larger than those with GPU DP and CPU DP operations when $dt \le 0.004$. But for $dt > 0.004$, there is no apparent difference in the curves of $\Delta E \sim dt$ with different numerical treatments due to that $dt$ has a dominant influence on relative energy RMS deviations in this range.
%For MD simulations with $r_{cut}= 6.0$, there is no apparent difference in the curves of $\Delta E \sim dt$ with different numerical treatments only when $dt > 0.004$. For $dt\le 0.004$, the RMS deviations of total energy in simulations with GPU SP and CPU SP operations are several times larger than those with GPU DP and CPU DP operations.
%We also found the shift function which can reduce the cutoff noise can significantly low the energy RMS deviations. Thereby, the standard GROMACS shift function~\cite{Lindahl2001} for LJ is provided by GALAMOST.
\par
%Since in CGMD simulations, the cutoff distance is normally quite small (for example $r_{cut}\le3.0$), and the integration time step is comparatively large (for example $dt\ge 0.002$), there should be no difference in simulation results by using GALAMOST with GPU DP and SP operations or by using GROMACS with CPU DP and SP operations.
Since in CGMD simulations, the cutoff distance is normally quite small (for example $r_{cut}\le3.0$), and the integration time step is comparatively large (for example $dt\ge 0.002$), the disturbance of numerical precision on MD course possesses a comparatively small weight. Thereby, there should be no any advantage of DP over SP in this type of simulation. However, we provide the GPU DP format of GALAMOST to cater to the need of high precision computation.
Due to different optimization strategies, the performances of GALAMOST with GPU DP operations for each model are different. We list the overall DP to SP performance ratios tested on a Tesla C2050 card in Table~\ref{performanceDP} for typical functions of GALAMOST.

%\section*{\sffamily \Large First-order heading}
%
%%((Equations should be inserted using standard LaTeX equation and eqnarray environments, not as graphics, and should be set in the main text))
%%Equation											(1)
%%((References should be superscripted and appear after punctuation.1,2 Please define all acronyms at their first usage except IR, UV, NMR, and DNA or similar commonly understood terms.))
%
%
%\subsection*{\sffamily \large Second-order heading}
%
%\subsubsection*{\sffamily \normalsize Third-order heading}
%
%{\sffamily \small Fourth-order heading}\\

%\section*{\sffamily \Large DISCUSSION}
%
%%((Place Discussion here. Not needed for review articles.))

\section*{\sffamily \Large CONCLUSIONS}

%((Place Conclusions here.))
This paper introduces the well-organized MD package GALAMOST with the incorporated force fields and characteristic models running on a single GPU. Except for common force fields, the MD-SCF method which can replace the intermolecular pair forces by the forces on particles interacting with density fields has been included in GALAMOST. For coarse-grained simulations of polymeric systems, a coarse-grained numerical potential method has also been incorporated in GALAMOST. The coarse-grained numerical potentials, including non-bonded, bond, angle, and torsion potentials, can be derived from IBI by fitting the reference structural distributions (such as RDF) to those from atomistic simulations or experiments.
\par
Besides these force fields, two characteristic models have been incorporated in GALAMOST. The model details and the algorithms of the soft anisotropic particle model which aims to simulate the self-assembled superstructures and other properties of anisotropic particles have been described. By developing new anisotropic potentials, novel anisotropic particles can be described by this model. In addition, we have introduced a chain-growth polymerization model and its implementation algorithm on the GPU. Many problems related to polymerization process can be studied by this model.
\par
The GALAMOST allows us to simulate larger polymeric systems over longer time. The highly-optimized algorithm for each method guarantees the efficiency of the implementation on the GPU. The characteristics of GALAMOST are embodied by the comparison of the performances for LJ liquid and DPD liquid with HOOMD-blue package. Specific performance of each characteristic method in GALAMOST has also been presented by benchmark comparisons with respect to corresponding CPU codes. The newly designed package GALAMOST with these lately developed MD techniques is dedicated to facilitate the studies of various polymeric systems.
\subsection*{\sffamily \large ACKNOWLEDGMENTS}
%((Place Acknowledgments here))
This work is subsidized by the National Basic Research Program of China (973 Program, 2012CB821500), and supported by National Science Foundation of China (21025416, 50930001).

%((Additional Supporting Information may be found in the online version of this article.))

\clearpage

%%%%%%%%%%%%%%%%%%%%%%%%%%%%%%%%%%%%%%%%%%%%%%%%%%%%%%%%%%%%%%%%%%%%%%%%%%%%%%%%%
% BIBLIOGRAPHY

%\bibliography{bibtexrefs}   % Produces the bibliography via BibTeX.
%\bibliography{galamost} %your .bib file
%\begin{thebibliography}{99}
%
%
%\bibitem{Coulson}
%Coulson, C. A., Rev. Mod. Phys., \textbf{1960}, 32,170-177.
%\bibitem{Malrieu}
%Malrieu, J.-P., J. Mol. Struct., \textbf{1998}, 424, 1-2,83-91.
%\bibitem{Shaik}
%Shaik, S., New. J. Chem., \textbf{2007}, 31,2015-2028.
%\bibitem{Hoffmann}
%Hoffman, R., Schleyer, P. v. R., Schaefer III, H. F., \textbf{2008}, 47, 7164-7167.
%\bibitem{Perdew}
%Perdew, J. P., Ruzsinszky, A., Constantin, L., Sun, J., Csonka, G., J. Chem. Theory Comput., \textbf{2009}, 5, 902-908.
%\bibitem{Koros}
%Koros, W. J.; Chern, R. T. In Handbook of Separation Process Technology; Rousseau, E. D.; Russell, B., Eds.; Wiley: New York, \textbf{1987}; Vol. 2, Chapter 20, pp 34-45.
%\end{thebibliography}

%%%%%%%%%%%%%%%%%%%%%%%%%%%%%%%%%%%%%%%%%%%%%%%%%%%%%%%%%%%%%%%%%%%%%%%%%%%%%%%%%

\clearpage
%%%%%%%%%%%%%%%%%%%%%%%%%%%%%%%%%%%%%%%%%%%%%%%%%%%%%%%%%%%%%%%%%%%%%%%%%%%%%%%%%
% FIGURE CAPTIONS

%%%%% FIGURE ---- cc.eps
\begin{figure}[!h]
\caption{\label{Fig:fig1} A column of Cell List table stores the tags or types together with coordinate vectors of the particles which are in the cell indicated by Cell Index. Cell Particle Number records the number of particles in each cell. A column of Neighbor List table stores the tags of the neighboring particles which are within the range of $r_{cut}$+$r_{buffer}$ of the particle indicated by Particle Tag. Neighbor Number records the number of the neighboring particles of each particle.}
\end{figure}
\begin{figure}[!h]
\caption{\label{Fig:fig2} A column of Bond List table stores the tags of the particles which connect with the particle indicated by Particle Tag. Bond Number records the number of the bonds of each particle.}
\end{figure}
\begin{figure}[!h]
\caption{\label{Fig:fig3} The structures of disk-like, rod-like, diblock Janus and triblock Janus particles from left side to right side, respectively.}
\end{figure}
\begin{figure}[!h]
\caption{\label{Fig:fig4} The average costing time per time step of GALAMOST and HOOMD simulating LJ liquid and DPD liquid systems with different system sizes both on GTX 580. These average values are measured after 1,000-10,000 time steps which depends on the system size. }
\end{figure}
%\begin{figure}[!h]
%\caption{\label{Fig:fig3} The performances for simulating triblock Janus systems by GALAMOST with different number of particles. The time steps per second (TPS) is measured on GeForce GTX 680.}
%\end{figure}
%\begin{figure}[!h]
%\caption{\label{Fig:fig4} The snapshot of triblock Janus system at 10 thousand time steps.}
%\end{figure}

\begin{figure}[!h]
\caption{\label{Fig:fig5} The comparison of electron density distributions of the components of DPPC and water at the state of equilibrium~\cite{Nicola2011} between OCCAM on CPU and GALAMOST on GeForce GTX 580. The simulations using grid size ($l$ = 0.587 nm, corresponding to 1.25$\sigma$) and update frequency of density filed (9 ps, corresponding to 300 time steps).}
\end{figure}
\begin{figure}[!h]
\caption{\label{Fig:fig6} The snapshots of DPPC and water systems with different contents of water at 20 million time steps. The systems A, B, C, and D correspond to the systems 3, 4, 5, and 6 in the Ref.~\cite{Nicola2012}, respectively. The colors of coarse-grained bead N, P, G, and C are blue, purple, red, and green, respectively. The simulations performed by GALAMOST are conducted on GeForce GTX 580. }
\end{figure}

\begin{figure}[!h]
\caption{\label{Fig:fig7} The comparison of the performances as million time steps/day between parallel CPU code OCCAM and GALAMOST in MD-SCF method about lipid and water systems LW1 (307,200 particles) and LW2 (1,048,576 particles) with a smaller grid size ($l=1.5\sigma$). The performances of OCCAM on parallel 96 CPUs are derived from Ref.~\cite{Zhao2012}, which were measured on cluster Crescol (Intel E7330, 2.4GHz). The performances of GALAMOST are measured on GeForce GTX 580.}
\end{figure}

\begin{figure}[!h]
\caption{\label{Fig:fig8} The comparisons of bond length and angle degree distributions for polystyrene melt between parallel CPU code IBIsCO and GALAMOST on GeForce GTX 680. More details about the polystyrene CG model can be found in Ref.~\cite{Qian2008}.}
\end{figure}

\begin{figure}[!h]
\caption{\label{Fig:fig9} The comparison of performances as TPS by simulating polystyrene melt with different number of particles between parallel CPU code IBIsCO~\cite{Karimi-Varzaneh2011} on Intel E5440, 2.83GHz and GALAMOST on GeForce GTX 680.}
\end{figure}

\begin{figure}[!h]
\caption{\label{Fig:fig10} A snapshot of surface initiated polymerization on the ball surface with ball  radius R$_{ball}=45$ is taken at 100 thousand time steps in simulation. The yellow particles which compose the surface of the ball are frozen. The blue particles on the surface of the ball are initiators and the green lines are the grafted polymers from initiators.}
\end{figure}
%\begin{figure}[!h]
%\caption{\label{Fig:fig11} The table of performances for polymerization systems with different radius of ball.}
%\end{figure}

\begin{figure}[!h]
\caption{\label{Fig:fig11} We have profiled the costing time of some time-consuming kernel functions running on the device for the system with R$_{ball}=15$ by using CUDA profiler. A: the costing time for once execution. B: the average costing time per time step. The neighbor list and the cell list are built once every 6.75 time steps and the polymerization function is executed once every 50 time steps in the simulation performed on Tesla C2050.}
\end{figure}

\begin{figure}[!h]
%\textcolor[rgb]{1.00,0.00,0.00}{
\caption{\label{Fig:fig12}  The RMS deviations of total energy $\Delta E$  are measured at different integration time steps with cutoff  distance $r_{cut}$ = 3.4 and $r_{cut}$ = 6.0 for the four types of numerical precisions. The value of each point in this figure is averaged from five parallel samples. }
\end{figure}

%%%%%%%%%%%%%%%%%%%%%%%%%%%%

%%%%%%%%%%%%%%%%%%%%%%%%%%%%%%%%%%%%%%%%%%%%%%%%%%%%%%%%%%%%%%%%%%%%%%%%%%%%%%%%%
% FIGURE FILES

\clearpage

%%\vspace*{0.1in}   %%% FIGURE 1
%\begin{center}
%\includegraphics[width=0.2\columnwidth,keepaspectratio=true]{cc.eps}
%\end{center}
%\vspace{0.25in}
%\hspace*{3in}
%{\Large
%\begin{minipage}[t]{3in}
%\baselineskip = .5\baselineskip
%Figure 1 \\
%Author A, Author B, Author C, Author D \\
%J.\ Comput.\ Chem.
%\end{minipage}
%}
%\newpage
\setcounter{figure}{0}
\newpage
\begin{figure}[!hbp]
\centering
\includegraphics[angle=0,width=0.95\textwidth]{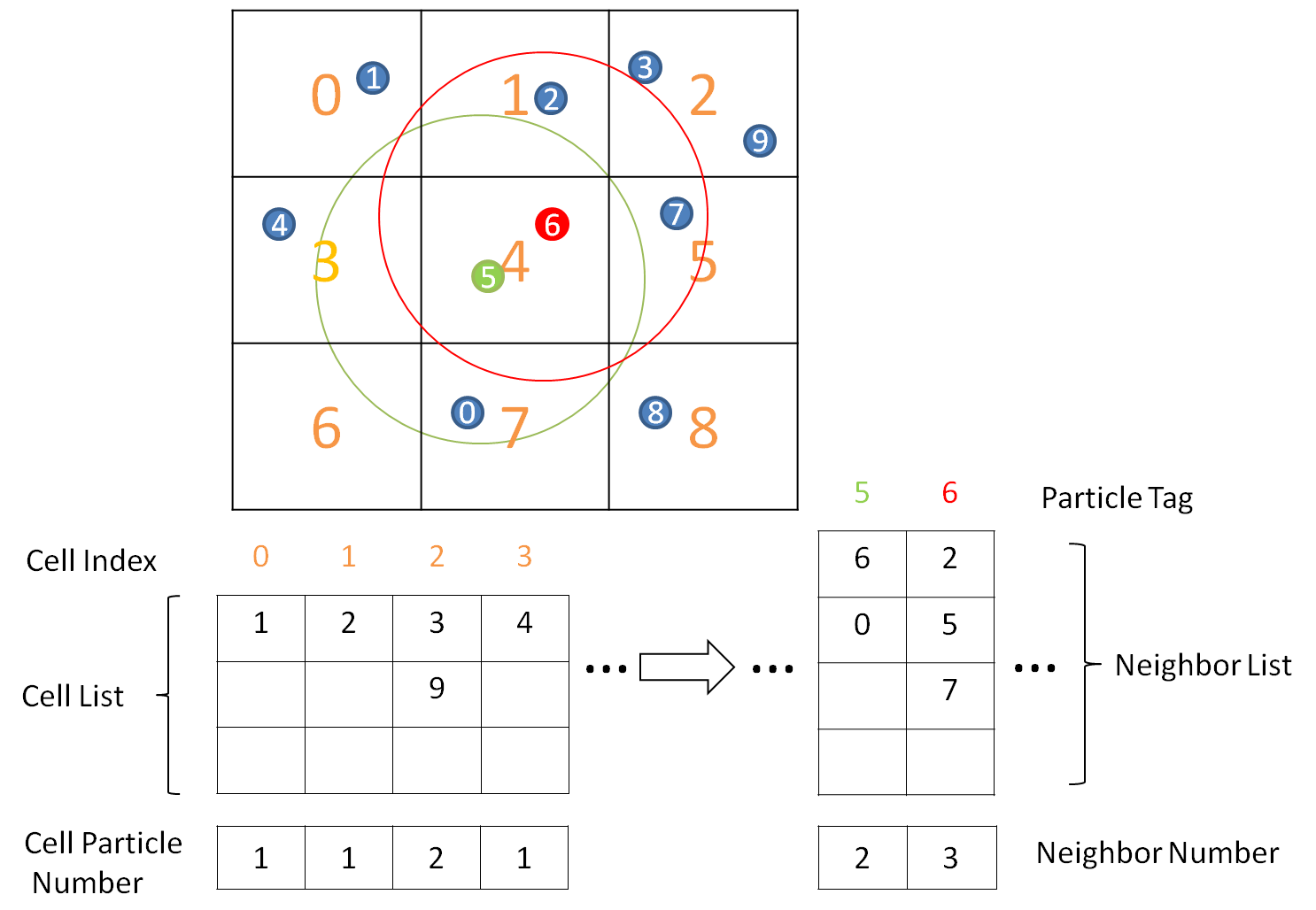}
\caption{Zhu et al. FIGURE 1}
\end{figure}

\newpage
\begin{figure}[!hbp]
\centering
\includegraphics[angle=0,width=0.95\textwidth]{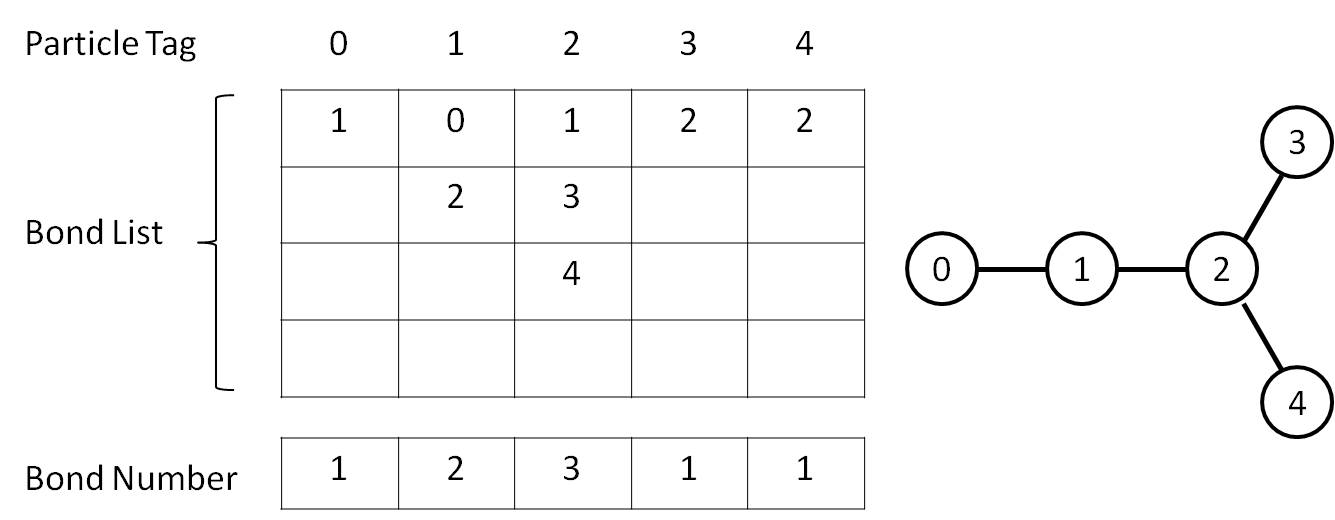}
\caption{Zhu et al. FIGURE 2}
\end{figure}

\newpage
\begin{figure}[!hbp]
\centering
\includegraphics[angle=0,width=0.95\textwidth]{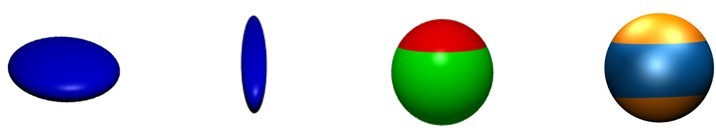}
\caption{Zhu et al. FIGURE 3}
\end{figure}

\newpage
\begin{figure}[!hbp]
\centering
\includegraphics[angle=0,width=0.95\textwidth]{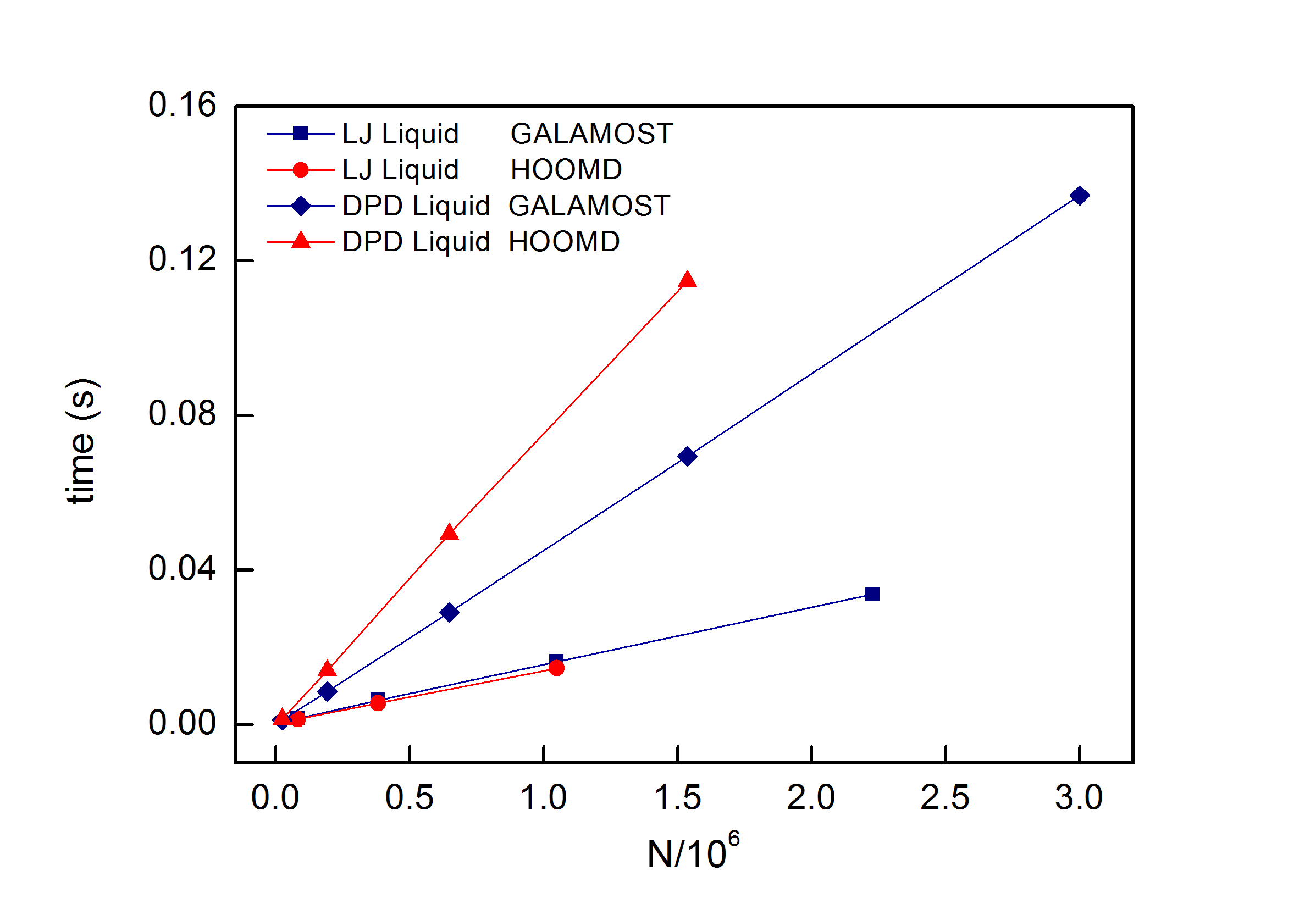}
\caption{Zhu et al. FIGURE 4}
\end{figure}

\newpage
\begin{figure}[!hbp]
\centering
\includegraphics[angle=0,width=0.95\textwidth]{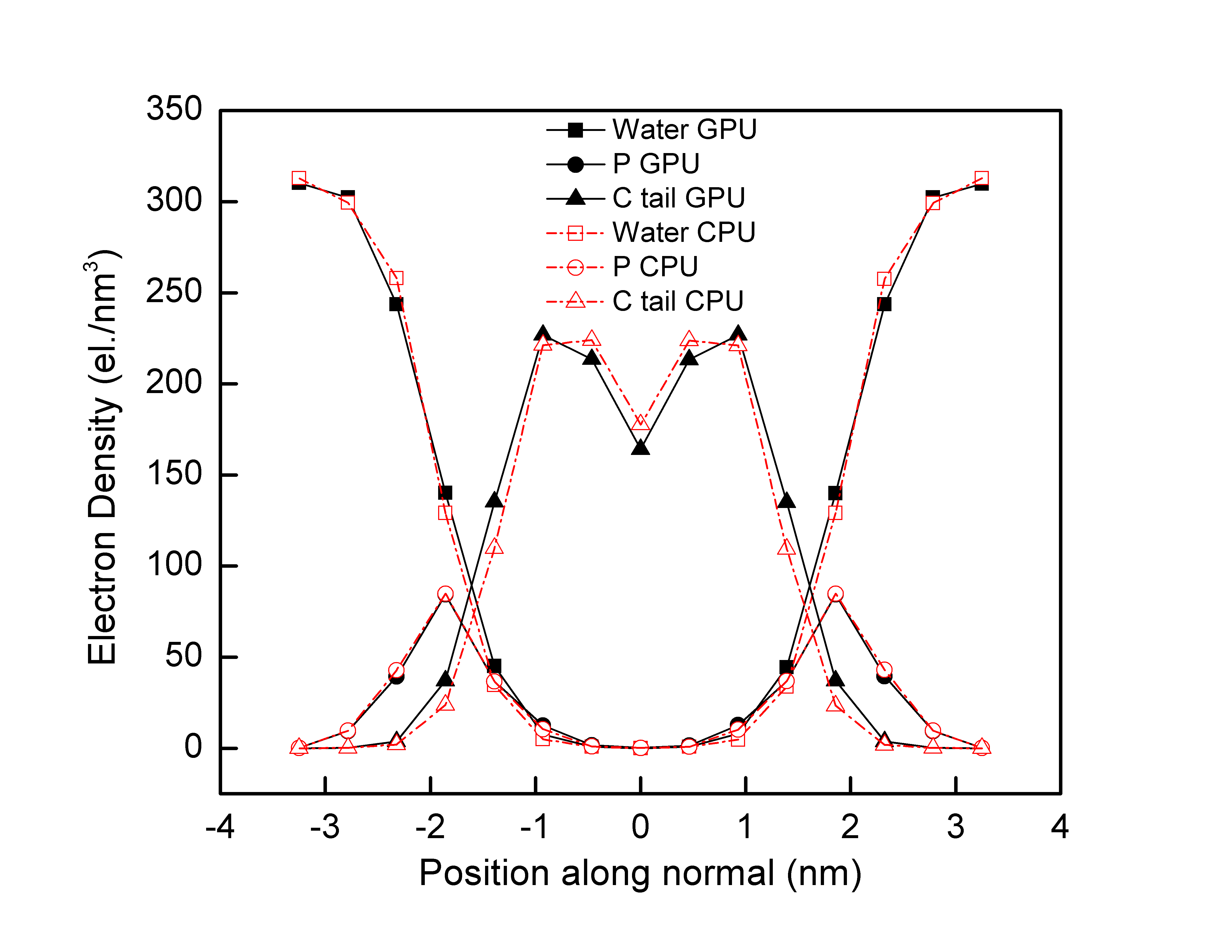}
\caption{Zhu et al. FIGURE 5}
\end{figure}

\newpage
\begin{figure}[!hbp]
\centering
\includegraphics[angle=0,width=0.95\textwidth]{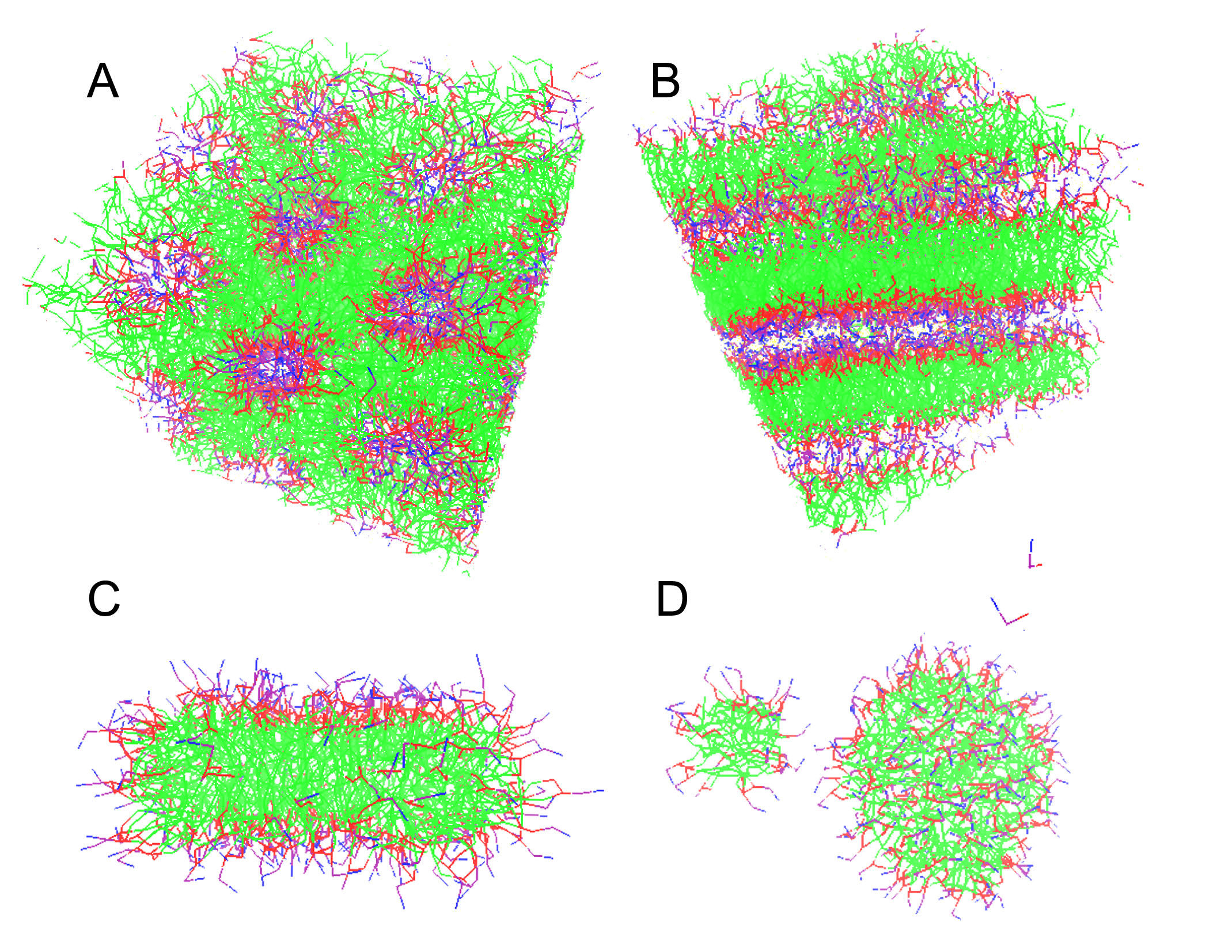}
\caption{Zhu et al. FIGURE 6}
\end{figure}

\newpage
\begin{figure}[!hbp]
\centering
\includegraphics[angle=0,width=0.95\textwidth]{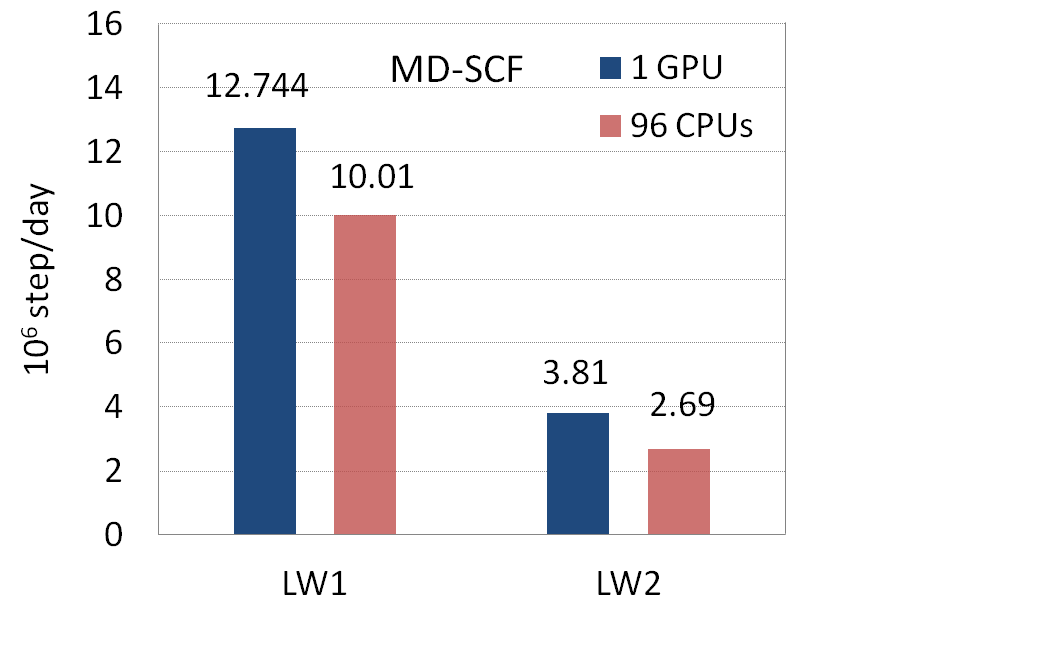}
\caption{Zhu et al. FIGURE 7}
\end{figure}

\newpage
\begin{figure}[!hbp]
\centering
\includegraphics[angle=0,width=0.95\textwidth]{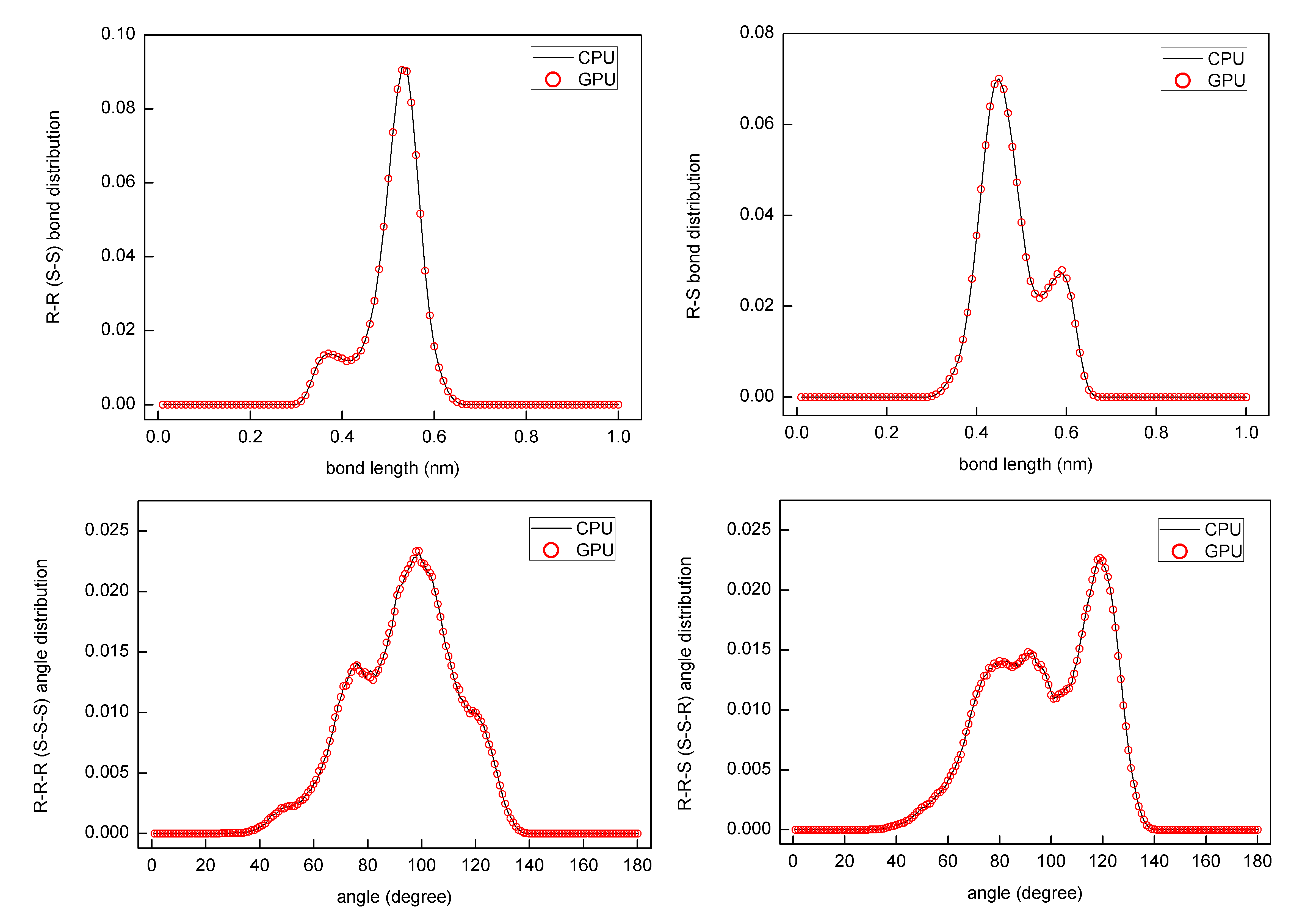}
\caption{Zhu et al. FIGURE 8}
\end{figure}

\newpage
\begin{figure}[!hbp]
\centering
\includegraphics[angle=0,width=0.95\textwidth]{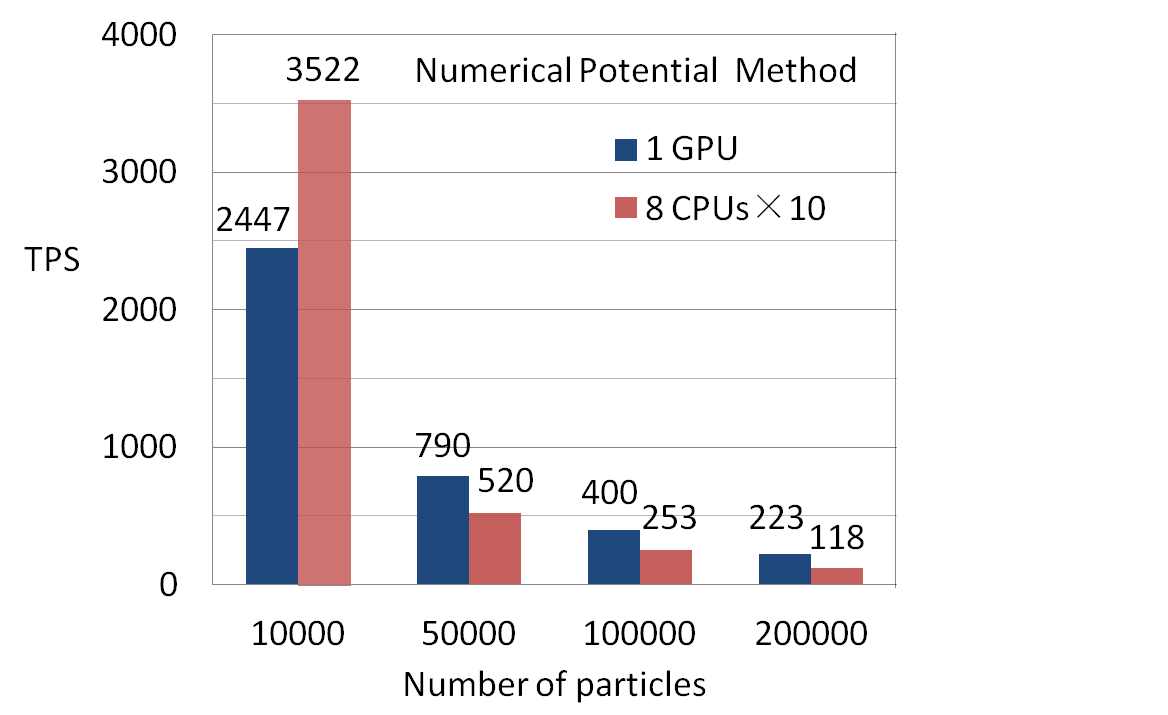}
\caption{Zhu et al. FIGURE 9}
\end{figure}

\newpage
\begin{figure}[!hbp]
\centering
\includegraphics[angle=0,width=0.95\textwidth]{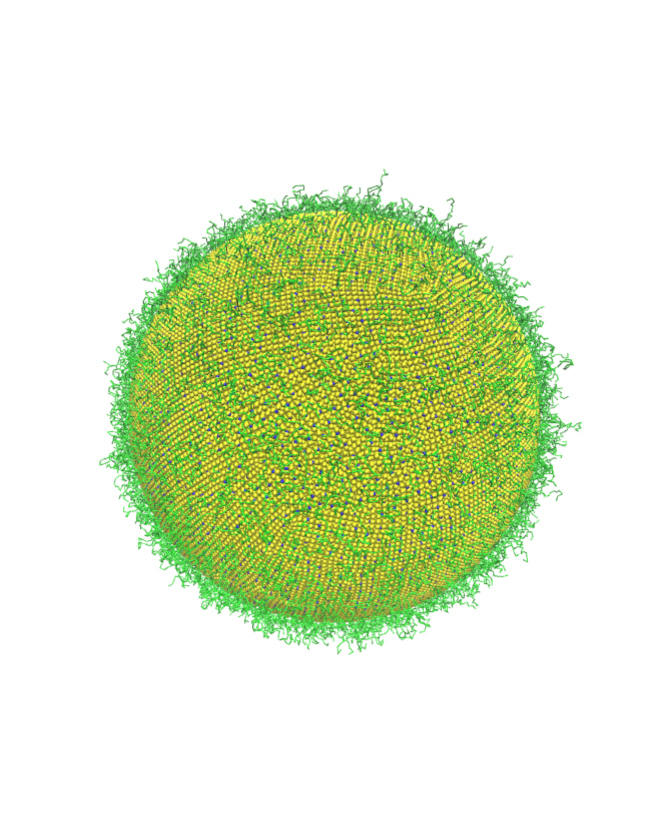}
\caption{Zhu et al. FIGURE 10}
\end{figure}

\newpage
\begin{figure}[!hbp]
\centering
\includegraphics[angle=0,width=0.95\textwidth]{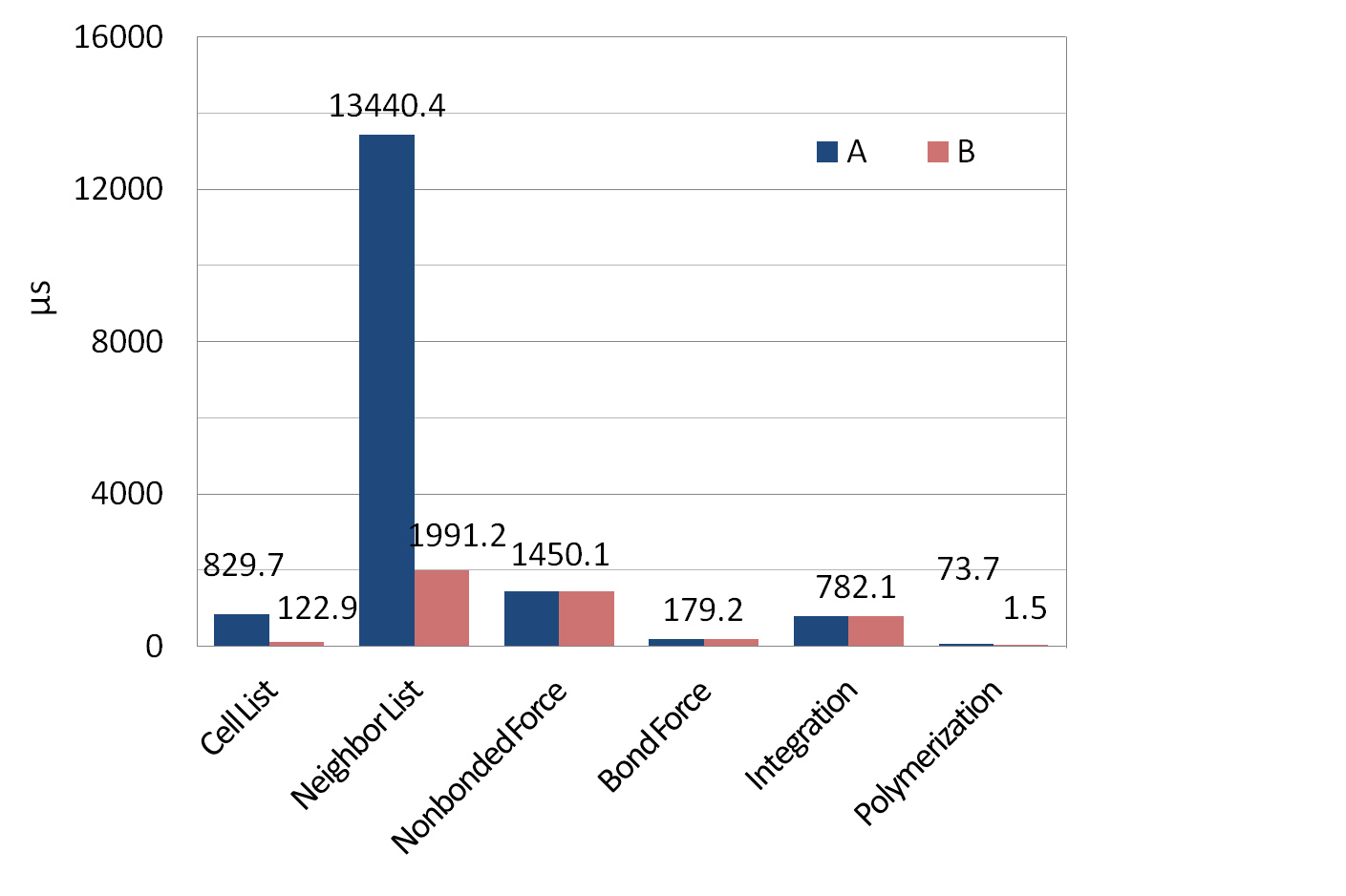}
\caption{Zhu et al. FIGURE 11}
\end{figure}

\newpage
\begin{figure}[!hbp]
\centering
\includegraphics[angle=0,width=0.95\textwidth]{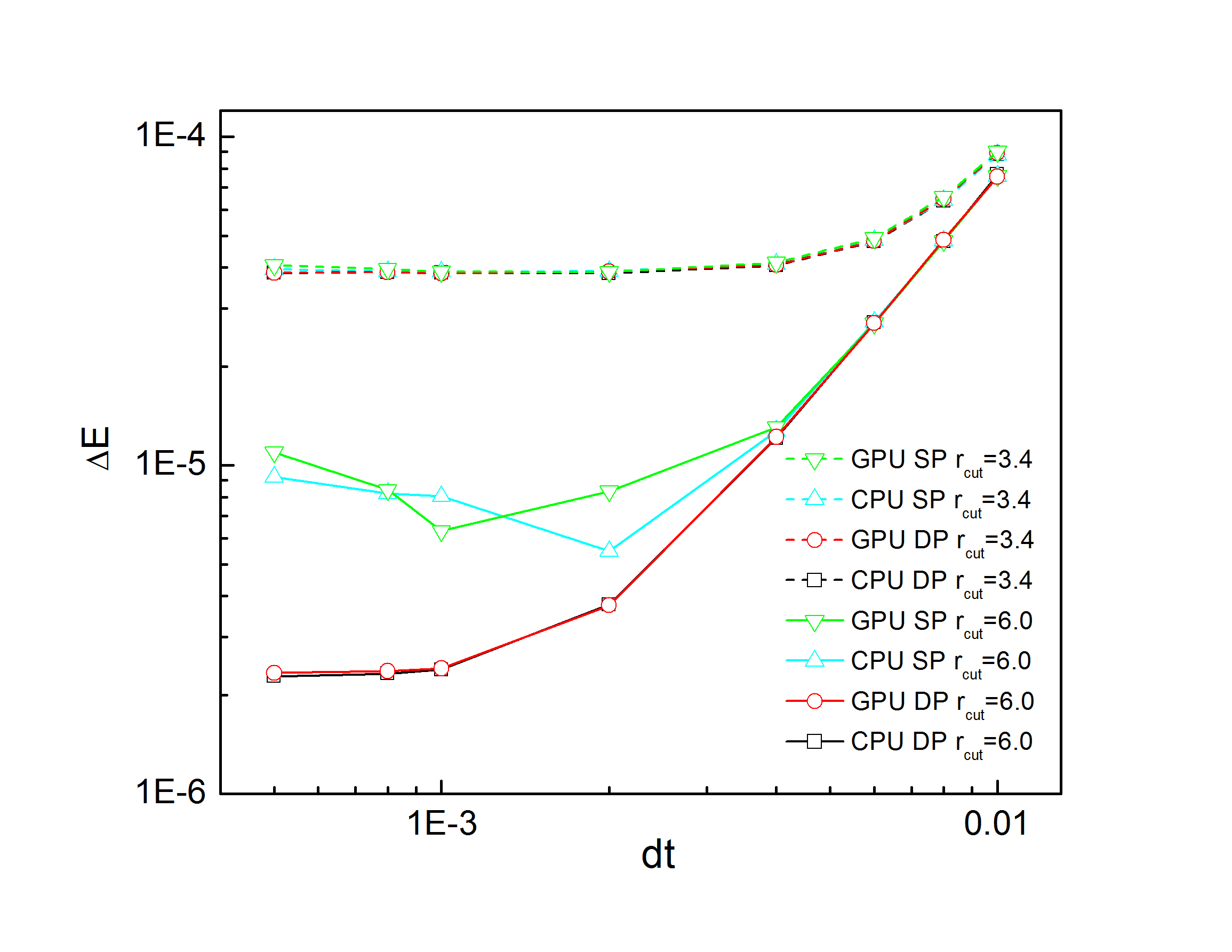}
\caption{Zhu et al. FIGURE 12}
\end{figure}
\clearpage

%\begin{table}
%\begin{tabular}{|c|c|c|c|}\hline
%\textbf{Quantity} & \textbf{Calculated} & \textbf{Observed} & \textbf{Error} \\ \hline
%  Density & 5.3 & 6.3 & Within limits \\ \hline
%  Optical magnification & 8.3 & 90.9 & Utterly unacceptable\! \\ \hline
%\end{tabular}
%\caption{\label{tbl1} Place table caption here.}
%\end{table}
\newpage
\begin{table}[!hbp]
\begin{tabular}{@{\qquad}l@{\qquad}c@{\qquad}c@{\qquad}c@{\qquad}c@{\qquad}c}\hline
\textbf{Algorithm 1} (Part 1) The intermolecular force calculation in MD-SCF\\
\hline
\textbf{Kernel 1} Building cell list\\
\textbf{Require}: (Np/blockDim) blocks run on the device\\
1. $i$$\leftarrow$blockIdx.x * blockDim.x + threadIdx.x\\
2. \textbf{if} $i$ $<$ Np \textbf{then}\\
3.{\qquad}$\vec{R}_{i}$$\Leftarrow$ $\vec{R}$; T$_{i}$$\Leftarrow$T\\
4.{\qquad}$cell\_index$$\leftarrow$($\vec{R}_{i}$)\\
5.{\qquad}CN$_{cell\_index}$$\Leftarrow$ atomicInc(\&CN[$cell\_index$], 0xffffffff) \\
6.{\qquad}$\vec{R}$$_{i}$, T$_{i}$$\Rightarrow$CL[($cell\_index$, CN$_{cell\_index}$)]\\
7. \textbf{end if}\\
\hline
\textbf{Kernel 2} Collecting density fractions at vertexes\\
\textbf{Require}: Nvert is the number of vertexes and equal to the number of cells \\
\textbf{Require}: D array records the density distribution at each vertex\\
\textbf{Require}: cell\_map array stores the indexes of eight neighboring cells of each vertex\\
\textbf{Require}: (Nvert/blockDim) blocks run on the device\\
1. $idx$$\leftarrow$blockIdx.x * blockDim.x + threadIdx.x\\
2. \textbf{if} $idx$ $<$Nvert \textbf{then}\\
3.{\qquad}\textbf{for} $k$= 0 \textbf{to} 8-1 \textbf{do}\\
4.{\qquad}{\qquad}$cell\_index$$\Leftarrow$cell\_map[$(idx, k)$]\\
5.{\qquad}{\qquad}\textbf{for} $j$=0 \textbf{to} CN[$cell\_index$]-1 \textbf{do}\\
6.{\qquad}{\qquad}{\qquad}$\vec{R}$$_{i}$, T$_{i}$$\Leftarrow$CL[$(cell\_index, j)$]\\
7.{\qquad}{\qquad}{\qquad}($\vec{R}$$_{i}$)+ D[$(idx,$ T$_{i}$)]$\Rightarrow$D[$(idx,$ T$_{i}$)]\\
8.{\qquad}{\qquad}\textbf{end for}\\
9.{\qquad}\textbf{end for}\\
10. \textbf{end if}\\
\hline
\end{tabular}
\end{table}

\newpage
\begin{table}[!hbp]
\begin{tabular}{@{\qquad}l@{\qquad}c@{\qquad}c@{\qquad}c@{\qquad}c@{\qquad}c}\hline
\textbf{Algorithm 1} (Part 2) The intermolecular force calculation in MD-SCF\\
\hline
\textbf{Kernel 3} Updating density gradients from density distributions\\
\textbf{Require}: NT is the number of particle types; $\vec{G}$ is density gradient vector array\\
\textbf{Require}: x, y ,and z are the indexes of a vertex in the three directions\\
\textbf{Require}: (Nvert/blockDim) blocks run on the device\\
1. $idx$$\leftarrow$blockIdx.x * blockDim.x + threadIdx.x\\
2. \textbf{if} $idx$ $<$Nvert \textbf{then}\\
3.{\qquad}x, y, z$\leftarrow(idx)$\\
4.{\qquad}\textbf{for} T =0 \textbf{to} NT \textbf{do}\\
5.{\qquad}{\qquad} D[((x,y,z), T)] - D[((x-1,y,z), T)]$\Rightarrow$$\vec{G}$[(idx, T, 1)].x\\
6.{\qquad}{\qquad} D[((x,y,z), T)] - D[((x,y-1,z), T)]$\Rightarrow$$\vec{G}$[(idx, T, 1)].y\\
7.{\qquad}{\qquad} D[((x,y,z), T)] - D[((x,y,z-1), T)]$\Rightarrow$$\vec{G}$[(idx, T, 1)].z\\
8.{\qquad}{\qquad} D[((x+1,y,z), T)] - D[((x,y,z), T)]$\Rightarrow$$\vec{G}$[(idx, T, 2)].x\\
9.{\qquad}{\qquad} D[((x,y+1,z), T)] - D[((x,y,z), T)]$\Rightarrow$$\vec{G}$[(idx, T, 2)].y\\
10.{\qquad}{\qquad}D[((x,y,z+1), T)] - D[((x,y,z), T)]$\Rightarrow$$\vec{G}$[(idx, T, 2)].z\\
11.{\qquad}\textbf{end for}\\
12. \textbf{end if}\\
\hline
\textbf{Kernel 4} Calculating intermolecular forces from density gradients\\
\textbf{Require}: (Np/blockDim) blocks run on the device\\
\textbf{Require}: xf1, yf1, zf1, xf2, yf2, zf2 are the fraction ratios of density gradients\\
1. $i$$\leftarrow$blockIdx.x * blockDim.x + threadIdx.x\\
2. \textbf{if} $i$ $<$Np \textbf{then}\\
3.{\qquad}$\vec{R}$$_{i}$$\Leftarrow$$\vec{R}$; T$_{i}$$\Leftarrow$T\\
4.{\qquad}x, y, z$\leftarrow$($\vec{R}_{i}$); idx$\leftarrow(x,y,z)$\\
5.{\qquad}xf1, yf1, zf1, xf2, yf2, zf2$\leftarrow$($\vec{R}$$_{i}$, x, y, z)\\
6.{\qquad}$\vec{F}_{i}$$\leftarrow$0 \\
7.{\qquad}\textbf{for} T$_{j}$ =0 \textbf{to} NT \textbf{do}\\
8.{\qquad}{\qquad} $\vec{F}_{i}$.x$\leftarrow$$\vec{F}_{i}$.x - $\chi_{T_{i},T_{j}}$$\ast$(xf1$\ast$$\vec{G}$[(idx, T$_{j}$, 1)].x+xf2$\ast$$\vec{G}$[(idx, T$_{j}$, 2)].x)\\
9.{\qquad}{\qquad} $\vec{F}_{i}$.y$\leftarrow$$\vec{F}_{i}$.y - $\chi_{T_{i},T_{j}}$$\ast$(yf1$\ast$$\vec{G}$[(idx, T$_{j}$, 1)].y+yf2$\ast$$\vec{G}$[(idx, T$_{j}$, 2)].y)\\
10.{\qquad}{\qquad}$\vec{F}_{i}$.z$\leftarrow$$\vec{F}_{i}$.z - $\chi_{T_{i},T_{j}}$$\ast$(zf1$\ast$$\vec{G}$[(idx, T$_{j}$, 1)].z+zf2$\ast$$\vec{G}$[(idx, T$_{j}$, 2)].z)\\
11.{\qquad}\textbf{end for}\\
12.{\qquad}$\vec{F}_{i}$ $\Rightarrow$$\vec{F}$\\
13. \textbf{end if}\\
\hline
\end{tabular}
\end{table}

\newpage
\begin{table}[!hbp]
\begin{tabular}{@{\qquad}l@{\qquad}c@{\qquad}c@{\qquad}c@{\qquad}c@{\qquad}c}\hline
\textbf{Algorithm 2} The non-bonded force calculation in numerical potential method \\
\hline
\textbf{Require}: (Np/blockDim) blocks run on the device\\
\textbf{Require}: Npoint is the number of the grid point of distance square\\
\textbf{Require}: c is the struct of c$_{0}$, c$_{1}$, c$_{2}$, and c$_{3}$\\
1. $i$$\leftarrow$blockIdx.x * blockDim.x + threadIdx.x\\
2. \textbf{if} $i$ $<$Np \textbf{then}\\
3.{\qquad}$\vec{R}_{i}$$\Leftarrow$$\vec{R}$; $T_{i}$$\Leftarrow$$T$; $\vec{F}_{i}$$\leftarrow$0\\
4.{\qquad}\textbf{for }$k$= 0 \textbf{to} NN[$i$]-1 \textbf{do}\\
5.{\qquad}{\qquad}$j$$\Leftarrow$NL[$(i,k)$]\\
6.{\qquad}{\qquad}$\vec{R}_{j}$$\Leftarrow$$\vec{R}$; $T_{j}$$\Leftarrow$$T$\\
7.{\qquad}{\qquad}$d\vec{r}$$\leftarrow$minimum image($\vec{R}_{i}$-$\vec{R}_{j}$)\\
8.{\qquad}{\qquad}$dr^{2}$$\leftarrow$$d\vec{r}$$\cdot$$d\vec{r}$\\
9.{\qquad}{\qquad}point$\leftarrow$int($dr^{2}$/interval)\\
10.{\qquad}{\qquad}\textbf{if} point $<$ Npoint \textbf{then}\\
11.{\qquad}{\qquad}{\qquad}c$_{ij}$$\Leftarrow$tex1Dfetch(c[(point,(T$_{i}$, T$_{j}$))])\\
12.{\qquad}{\qquad}{\qquad}$\delta$$\leftarrow$$dr^{2}$ - float(point)$\ast$interval \\
13.{\qquad}{\qquad}{\qquad}$\vec{F}_{i}$$\leftarrow$$\vec{F}_{i}$ + (c$_{ij}$, $\delta$) \\
14.{\qquad}{\qquad}\textbf{end if}\\
15.{\qquad}\textbf{end for}\\
16.{\qquad}$\vec{F}_{i}$$\Rightarrow$$\vec{F}$\\
17. \textbf{end if}\\
\hline
\end{tabular}
\end{table}

\newpage
\begin{table}[!hbp]
\begin{tabular}{@{\qquad}l@{\qquad}c@{\qquad}c@{\qquad}c@{\qquad}c@{\qquad}c}\hline
\textbf{Algorithm 3} Anisotropic particle model with Berendsen thermostat\\
\hline
\textbf{Kernel 1} First step integration\\
\textbf{Require}: (Np/blockDim) blocks run on the device\\
1. $i$$\leftarrow$blockIdx.x * blockDim.x + threadIdx.x\\
2. \textbf{if} $i$ $<$ Np \textbf{then}\\
3.{\qquad}$\vec{R}_{i}$$\Leftarrow$ $\vec{R}_{t-dt}$; $\vec{V}_{i}$$\Leftarrow$ $\vec{V}_{t-dt/2}$\\
4.{\qquad}$\vec{n}_{i}$$\Leftarrow$ $\vec{n}_{t-dt}$; $\vec{u}_{i}$$\Leftarrow$ $\vec{u}_{t-dt/2}$\\
5.{\qquad}$\vec{R}_{i}$+$\vec{V}_{i}$$\ast$dt$\Rightarrow$$\vec{R}_{t}$  \\
6.{\qquad}$\vec{n}_{i}$+$\vec{u}_{i}$$\ast$dt$\Rightarrow$$\vec{n}_{t}$  \\
7. \textbf{end if}\\
\hline
\textbf{Kernel 2} Calculating force and torque\\
\textbf{Require}: (Np/blockDim) blocks run on the device\\
1. $i$$\leftarrow$blockIdx.x * blockDim.x + threadIdx.x\\
2. \textbf{if} $i$ $<$Np \textbf{then}\\
3.{\qquad}$\vec{R}_{i}$$\Leftarrow$ $\vec{R}_{t}$; $\vec{n}_{i}$$\Leftarrow$ $\vec{n}_{t}$; $\vec{F}_{i}$$\leftarrow$0; $\vec{g}_{i}$$\leftarrow$0\\
4.{\qquad}\textbf{for }$k$= 0 \textbf{to} NN[$i$]-1 \textbf{do}\\
5.{\qquad}{\qquad}$j$$\Leftarrow$NL[$(i, k)$]\\
6.{\qquad}{\qquad}$\vec{R}_{j}$$\Leftarrow$ $\vec{R}_{t}$; $\vec{n}_{j}$$\Leftarrow$ $\vec{n}_{t}$\\
7.{\qquad}{\qquad}$\vec{F}_{i}$$\leftarrow$$\vec{F}_{i}$+($\vec{R}_{i}$,$\vec{R}_{j}$,$\vec{n}_{i}$,$\vec{n}_{j}$)\\
8.{\qquad}{\qquad}$\vec{g}_{i}$$\leftarrow$$\vec{g}_{i}$+($\vec{R}_{i}$,$\vec{R}_{j}$,$\vec{n}_{i}$,$\vec{n}_{j}$)\\
9.{\qquad}\textbf{end for}\\
10.{\qquad}$\vec{F}_{i}$$\Rightarrow$$\vec{F}_{t}$; $\vec{g}_{i}$-($\vec{g}_{i}$$\cdot$$\vec{n}_{i}$)$\vec{n}_{i}$$\Rightarrow$$\vec{g}_{t}^{\perp}$\\
11. \textbf{end if}\\
\hline
\textbf{Kernel 3} Second step integration\\
\textbf{Require}: (Np/blockDim) blocks run on the device\\
\textbf{Require}: ST and SRT are scaling factors to control\\
{\qquad}{\qquad}  translational and rotational temperature, respectively\\
1. $i$$\leftarrow$blockIdx.x * blockDim.x + threadIdx.x\\
2. \textbf{if} $i$ $<$ Np \textbf{then}\\
3.{\qquad}$\vec{F}_{i}$$\Leftarrow$ $\vec{F}_{t}$; $\vec{g}_{i}^{\perp}$$\Leftarrow$ $\vec{g}_{t}^{\perp}$; $\vec{n}_{i}$$\Leftarrow$ $\vec{n}_{t}$\\
4.{\qquad}$\vec{u}_{i}$$\Leftarrow$ $\vec{u}_{t-dt/2}$; $\vec{V}_{i}$$\Leftarrow$ $\vec{V}_{t-dt/2}$\\
5.{\qquad}$\vec{u}_{i}$+(($\vec{g}_{i}^{\perp}$/inertia$_{i}$)$\ast$dt - $\lambda$($\vec{u}_{i}$$\cdot$$\vec{n}_{i}$)$\vec{n}_{i}$)$\ast$SRT$\Rightarrow$$\vec{u}_{t+dt/2}$ \\
6.{\qquad}$\vec{V}_{i}$+($\vec{F}_{i}$/mass$_{i}$)$\ast$dt$\ast$ST$\Rightarrow$$\vec{V}_{t+dt/2}$  \\
7. \textbf{end if}\\
\hline
\end{tabular}
\end{table}

\newpage
\begin{table}[!hbp]
\begin{tabular}{@{\qquad}l@{\qquad}c@{\qquad}c@{\qquad}c@{\qquad}c@{\qquad}c}\hline
\textbf{Algorithm 4} Chain-growth polymerization model\\
\hline
\textbf{Require}: N$_{init}$ is the number of initiators, (N$_{init}$/blockDim) blocks run on the device\\
\textbf{Require}: Group$_{init}$ stores the tags of initiators\\
\textbf{Require}: Init array indicates each particle if it is an initiator: 1 true, 0 false \\
\textbf{Require}: Cris array indicates each particle if it is a monomer: 1 false, 0 true \\
\textbf{Require}: Random(0,1) is a pseudo random number generator and generates \\
{\qquad}{\qquad}  float numbers in a range from 0 to 1\\
1. $idx$$\leftarrow$blockIdx.x * blockDim.x + threadIdx.x\\
2. \textbf{if} $idx$ $<$ N$_{init}$ \textbf{then}\\
3.{\qquad}$i$$\Leftarrow$ Group$_{init}$[$idx$]\\
4.{\qquad}$\vec{R}_{i}$$\Leftarrow$ $\vec{R}$[$i$]   \\
5.{\qquad}\textbf{for} $k$= 0 \textbf{to} NN[$i$] - 1\textbf{do}\\
6.{\qquad}{\qquad}$j$$\Leftarrow$NL[$(i, k)$] \\
7.{\qquad}{\qquad}$\vec{R}_{j}$$\Leftarrow$ $\vec{R}$[$j$] \\
8.{\qquad}{\qquad}I$_{j}$ $\Leftarrow$ Init[$j$]; C$_{j}$ $\Leftarrow$ Cris[$j$]\\
9.{\qquad}{\qquad}$d\vec{r}$ $\leftarrow$ minimum image($\vec{R}_{i} -\vec{R}_{j}$) \\
10.{\qquad}{\qquad}\textbf{if} $|d\vec{r}|<r_{cut}$ \textbf{and} C$_{j}$==0 \textbf{and} I$_{j}$==0 \textbf{then}\\
11.{\qquad}{\qquad}{\qquad}\textbf{if} Random(0,1)$<$ Pr$_{ij}$ \textbf{then}\\
12.{\qquad}{\qquad}{\qquad}{\qquad}oldValue $\Leftarrow$ atomicMax(\&Cris[$j$],1)\\
13.{\qquad}{\qquad}{\qquad}{\qquad}\textbf{if} oldValue == 0 \textbf{then}\\
14.{\qquad}{\qquad}{\qquad}{\qquad}{\qquad}$j$$\Rightarrow$Group$_{init}$[$idx$]\\
15.{\qquad}{\qquad}{\qquad}{\qquad}{\qquad}$1\Rightarrow$Init[$j$]; 0$\Rightarrow$Init[$i$]; 1$\Rightarrow$Cris[$i$]\\
16.{\qquad}{\qquad}{\qquad}{\qquad}{\qquad}Num$_{i}$$\Leftarrow$BN[$i$]; Num$_{j}$$\Leftarrow$BN[$j$]\\
17.{\qquad}{\qquad}{\qquad}{\qquad}{\qquad}$j$$\Rightarrow$BL[($i$ ,Num$_{i}$)]; i$\Rightarrow$BL[($j$, Num$_{j}$)]\\
18.{\qquad}{\qquad}{\qquad}{\qquad}{\qquad}Num$_{i}$+1$\Rightarrow$BN[$i$]; Num$_{j}$+1$\Rightarrow$BN[$j$]\\
19.{\qquad}{\qquad}{\qquad}{\qquad}{\qquad}\textbf{break}\\
20.{\qquad}{\qquad}{\qquad}{\qquad}\textbf{end if}\\
21.{\qquad}{\qquad}{\qquad}\textbf{end if}\\
22.{\qquad}{\qquad}\textbf{end if}\\
23.{\qquad}\textbf{end for}\\
24. \textbf{end if}\\
\hline
\end{tabular}
\end{table}

\newpage
\begin{table}[!hbp]
\caption{The performances of GALAMOST with anisotropic particle model simulating triblock Janus systems with different number of particles. The time steps per second (TPS) are measured on GeForce GTX 680. }\label{triblockJanus} \centering
\begin{tabular}{|@{\quad}c@{\quad}|@{\quad}c@{\quad}|@{\quad}c@{\quad}|@{\quad} c@{\quad}|@{\quad}c@{\quad}|}
\hline
Box size&(20)$^{3}$&(40)$^{3}$&(60)$^{3}$&(80)$^{3}$\\
\hline
Number of particles&24,000&192,000&648,000&1,536,000\\
\hline
TPS& 945.4$\pm$0.9 & 128.2$\pm$0.2 &37.1$\pm$0.1&15.3$\pm$0.04\\
\hline
\end{tabular}
\end{table}

\begin{table}[!hbp]
\caption{The performances are measured on Tesla C2050 of GALAMOST with chain-growth polymerization model simulating the surface-initiated polymerization with different system sizes. $\Sigma$ is initiator density which indicates the number of initiators per unit area. R$_{ball}$ is the radius of the ball. N$_{init}$ is the number of initiators. N$_{frozen}$ is the number of frozen particles on the surface of the ball. N$_{free}$ is the number of monomers. N$_{total}$ is the number of all particles.}\label{polymerization}\centering
\begin{tabular}{|@{\quad}c@{\quad}|@{\quad}c@{\quad}|@{\quad}c@{\quad}|@{\quad} c@{\quad}|@{\quad}c@{\quad}|@{\quad}c@{\quad}|@{\quad}c@{\quad}|@{\quad}c@{\quad}|}
\hline
R$_{ball}$&$\Sigma$&N$_{init}$&N$_{frozen}$&N$_{free}$&N$_{total}$&Box Size&TPS\\
\hline
5&0.528&166&1,086&47,124&48,210&(38.25)$^{3}$&780.9$\pm$1.2\\
\hline
10&0.508&639&4,381&18,8495&192,876&(60.91)$^{3}$&220.4$\pm$0.4\\
\hline
15&0.506&1,431&9,932&42,4114&434,046&(80.06)$^{3}$&126.4$\pm$0.3\\
\hline
25&0.507&3,982&28,120&1,178,096&1,206,216&(113.22)$^{3}$&43.28$\pm$0.09\\
\hline
45&0.507&1,2926&90,561&3,817,032&3,907,593&(169.53)$^{3}$&13.31$\pm$0.07\\
\hline
\end{tabular}
\end{table}

\begin{table}[!hbp]
%\textcolor[rgb]{1.00,0.00,0.00}{
\caption{The performances of GALAMOST in GPU DP and GPU SP operations are measured on the same Tesla C2050 card. The DP to SP performance ratios for each method and model are obtained for the systems with particle number ranging from $2\times10^{4}$ to $4\times10^{5}$. }\label{performanceDP}\centering
\begin{tabular}{|@{\quad}c@{\quad}|@{\quad}c@{\quad}|@{\quad}c@{\quad}|}
\hline
Method and model&System&DP/SP ratio\\
\hline
MD-SCF&DPPC/water&0.56-0.63\\
\hline
Numerical potential method&polystyrene melt&0.34-0.38\\
\hline
Anisotropic particle model&Janus particle&0.36-0.40\\
\hline
Polymerization model&surface polymerization&0.45-0.47\\
\hline
\end{tabular}
\end{table}
\end{document}